\documentclass[aps,twocolumn,prd]{revtex4}

\usepackage{amssymb}
\usepackage{amsmath}
\usepackage{graphicx}
\usepackage[dvips]{color}
 
\newcommand{\ar}{\arrowvert}

\newcommand{\da}{\dagger} 
\newcommand{\ov}{\overline} 
 
\newcommand{\be}{\begin{equation}} 
\newcommand{\ee}{\end{equation}} 
\newcommand{\ba}{\begin{eqnarray}} 
\newcommand{\ea}{\end{eqnarray}}

\newcommand{\pa}{\partial} 

\begin{document} 
\title{Transport properties of bottomed mesons in a hot mesonic gas}
               
\author{ Luciano M. Abreu$^1$, Daniel Cabrera$ ^2$, and Juan M. Torres-Rincon$
^3$ }
 
\affiliation{ $^{1}$Instituto de F{\'i}sica, Universidade Federal da
Bahia, 40210-340, Salvador, BA, Brazil}
\affiliation{ $ ^2$ Departamento de F\'{\i}sica Te\'orica II,  Universidad 
Complutense, 28040 Madrid, Spain}
\affiliation{ $^3$ Departamento de F\'{\i}sica Te\'orica I,  Universidad 
Complutense, 28040 Madrid, Spain }

\begin{abstract}
In this work we evaluate the $B$-meson drag and diffusion coefficients in a hot
medium constituted of light mesons ($\pi$, $K$, $\bar{K}$ and $\eta$). We treat
the $B$-meson and $B^*$-meson interaction with pseudo-Goldstone bosons in chiral
perturbation theory at next-to-leading order within the constraints from heavy
quark symmetry, and employ standard unitarization techniques of NLO amplitudes
in order to account for dynamically generated resonances (leading to a more
efficient heavy-flavor diffusion) and thus reach higher temperatures. We
estimate individual meson contributions from the gas to the transport
coefficients and perform a comparison with other findings in literature. We
report a bottom relaxation length of about 80~fm at a temperature of 150 MeV and for typical momenta of 1~GeV,  at
which our approach is reliable. Compared to a charm relaxation length of 40 fm
in the same conditions, we conclude that the $B$ mesons provide a cleaner probe
of the early stages of a heavy-ion collision. 
\end{abstract} 

\maketitle

\section{Introduction} 

The features of matter formed in heavy ion collisions (HICs) have been a subject
of great interest in the last decades. In this scenario, heavy-flavored hadrons
play an essential role since they carry heavy quarks produced in the early stage
of the collisions. Therefore, heavy mesons are interesting probes to understand
the evolution of partonic matter since its creation, unlike pions and kaons which
can be produced in the thermal medium at later stages. 

However, it is worth noticing that the momentum spectra of charmed and bottomed
mesons extracted from HICs undergo modifications due to their interactions with
the hadron medium, constituted of pions and other particles. In this sense, the
diffusion of heavy mesons in an equilibrium hadronic gas must be taken into
account, and may be properly studied in the framework of kinetic theory to
compute transport coefficients.

Different approaches have been used to study various aspects of this topic~
\cite{Laine:2011is,He:2011yi,Ghosh:2011bw,LADCFLEJT:2011,He:2011qa,Das:2011}. In
particular, in Ref.~\cite{Laine:2011is} heavy quark effective theory (HQET) and
chiral perturbation theory (ChPT) have been employed, focusing on the lowest
possible temperatures. Attempts to reach higher temperatures in the context of
charmed mesons close to the crossover to the quark and gluon plasma have been
done in Refs.~\cite{He:2011yi,Ghosh:2011bw}. In addition, in
Ref.~\cite{Das:2011} the transport coefficients of $B$ mesons have been obtained
with the use of scattering lengths as dynamical input.

In our recent work \cite{LADCFLEJT:2011} the transport coefficients of charmed
mesons in a hot pion gas were computed, exploiting ChPT at next-to-leading order
(NLO) and employing standard unitarization as guiding principle to reach higher
temperatures and account for the contribution of resonant channels.
Thus, a natural question arises about the application of this
approach to bottomed mesons, which would allow a comparison with existing
literature and, hence, a better comprehension of this issue. 

In the present work we extend the framework used in \cite{LADCFLEJT:2011} to
evaluate the $B$-meson drag and diffusion coefficients in a hot mesonic gas,
including pions, kaons and $\eta$ mesons, with special attention to the
contribution of the heavier states with respect to the pion gas. We perform a
detailed analysis of the temperature and momentum dependence of these
coefficients as well as their static limit (vanishing heavy-meson momentum) and scaling
properties with the heavy-meson mass.

The organization of this paper is as follows. In Sec.~\ref{sec:formalism}, we
introduce the transport coefficients and the chiral Lagrangian density that
describes the interactions between $B$ mesons and light mesons. Afterwards,
Sec.~\ref{sec:scatt_matrix} is devoted to obtain the unitarized scattering
amplitude and fit the relevant free constants to available data. The transport
coefficients are evaluated and analyzed in detail in Sec.~\ref{sec:transport}
for a pure pion gas. A pertinent discussion of the role of unitarization of
scattering amplitudes in heavy-quark diffusion is also contained in this
section. The  modifications in the transport coefficients due to inclusion of
kaons and $\eta$ mesons in the thermal bath are studied in Sec.~\ref{sec:kaons}. A
summary and concluding remarks are given in Sec.~\ref{sec:summary}.

\section{\label{sec:formalism} Formalism}

\subsection{Transport coefficients} 

We evaluate the transport coefficients of stable $B$ mesons propagating through a hot
mesonic gas by using the scattering amplitudes obtained from chiral perturbation
theory, taking into account unitarity and heavy-quark symmetry. We assume that
the density of pseudoscalar ($B$) and vector ($B^*$) bottomed mesons is very
small, so we will neglect collisions among bottomed mesons themselves and
concentrate only on their interaction with the light meson gas in thermal
equilibrium. 

Following the assumptions made in Ref.~\cite{LADCFLEJT:2011}, the momentum-space
distribution of  bottomed mesons must relax via the Fokker-Planck equation. We
choose the momenta of the elastic collision between a meson $B^{(*)}$ and a
light meson $\phi$ as
\be B ^{(*)}(\mathbf{p}) + \phi (\mathbf{q}) \rightarrow
B^{(*)}(\mathbf{p}-\mathbf{k}) + \phi (\mathbf{q} +\mathbf{k}) \ . \ee
In this context, the evolution of the momentum distribution of bottomed mesons
due to their
interaction with the isotropic mesonic gas is fully controlled by the drag ($F$) and
diffusion ($\Gamma _0,\Gamma_1$) coefficients, written as 
\begin{eqnarray} \label{Transportintegrals}
 F(p^2) & = & \int d\mathbf{k}\  w(\mathbf{p},\mathbf{k})  \ \frac{k_ip^i}{p^2} \ , \\ \nonumber
 \Gamma_0 (p^2) & = &  \frac{1}{4} \int d\mathbf{k}\ w(\mathbf{p},\mathbf{k}) \left[ \mathbf{k}^2 - \frac{(k_i p^i)^2}{p^2} \right] \ , \\ \nonumber
 \Gamma_1(p^2) & = &  \frac{1}{2} \int d\mathbf{k}\  w(\mathbf{p},\mathbf{k}) \
 \frac{(k_i p^i)^2}{p^2} \ ,
\end{eqnarray}
where $w (\mathbf{p}, \mathbf{k} )$ denotes the collision rate for a bottomed meson
with initial (final) momentum $\mathbf{p}$ ($\mathbf{p}-\mathbf{k}$), 
\begin{widetext}
\be 
w (\mathbf{p}, \mathbf{k} ) = g_{\phi} \int \frac{d \mathbf{q}}{(2\pi)^9} f_{\phi} (\mathbf{q}) \left[ 1+ f_{\phi} (\mathbf{q}+\mathbf{k}) \right]
  \frac{1}{2E_p^B}  \frac{1}{2E_q^{\phi}} \frac{1}{2 E_{p-k}^B} \frac{1}{2 E_{q+k}^{\phi}}
 (2\pi)^4 \delta (E_p^B + E_q^{\phi} - E^B_{p-k} -E_{q+k}^{\phi} )  \sum |\mathcal{M}_{B \phi }(s,t,\chi)|^2 . 
 \label{probdist}
\ee
\end{widetext}
\noindent
In Eq. (\ref{probdist}), $f_{\phi} (\mathbf{q})$ is the bath's distribution
function; $\mathcal{M}_{B \phi }$ stands for the Lorentz invariant $B$ meson -
light meson 
scattering matrix element, $g_\phi$ is the Goldstone boson isospin
degeneracy (i.e. $g_\pi=3$ for the pion), and $\chi$ denotes the possible spin
degrees of freedom. 

In the next subsection we derive the scattering amplitude $\mathcal{M}$ for 
bottomed mesons  in the light meson medium, necessary to evaluate the three
transport coefficients introduced above.

\subsection{Effective Lagrangian for $B$-meson and light meson interaction} 

Our task now is to construct the chiral Lagrangian density that describes the
interactions between the $J=0$ and $J=1$ $B$ mesons and light mesons. In this sense, we note that heavy-meson ChPT, described in Refs.~\cite{Lutz:2007sk,Guo:2009ct,Geng:2010vw,LADCFLEJT:2011} for the case of charmed mesons, can be applied to the $B$-meson sector as well.

Let us start by introducing the pseudoscalar Goldstone bosons.
They follow the nonlinear realization of the $SU(N_f)_L \times SU(N_f)_R$ chiral symmetry, given in the exponential parametrization
\be U = \exp{\left( \frac{\sqrt{2}i \phi}{F} \right)} \ , \ee
with $F$ being the Goldstone boson decay constant in the chiral limit and $\phi$
the matrix incorporating the pseudoscalar Goldstone bosons,   
\be
\phi = \left( \begin{array}{ccc}
  \frac{1}{\sqrt{2}}\pi ^{0}+\frac{1}{\sqrt{6}} \eta  & \pi ^{+} & K^{+} \\
    \pi^{-} & -\frac{1}{\sqrt{2}}\pi ^{0}+\frac{1}{\sqrt{6}} \eta & K^0 \\
    K^{-}   & \bar{K}^{0}  &  -\frac{2}{\sqrt{6}} \eta
\end{array}
            \right).
\label{phi}
\ee
Under chiral symmetry, the matrix $U$ is transformed as
\be U \rightarrow U' = L U R^{\dag} \ , \ee
where $L$ and $R$ are global transformations under $SU(3)_L$ and $SU(3)_R$, respectively.
The Goldstone boson kinetic term of the effective Lagrangian is explicitly invariant under this symmetry
\be \mathcal{L}^{\phi} = \frac{F^2}{4}  \langle \pa_\mu U^\dag \pa^\mu U \rangle \ . \ee
For convenience, a matrix $u$ is introduced as
\be u = \sqrt{U} \ , \ee
and is transformed under chiral symmetry as
\be u \rightarrow u' = Lu W^\dag =W u R^{\dag} \ , \ee
where $W$ is a unitary matrix expressible as a certain combination of $L$, $R$ and $\phi$.
The axial and vector fields are constructed as
\ba
\Gamma _{\mu} & = & \frac{1}{2} \left( u^{\da} \partial _{\mu} u +  u\partial _{\mu} u^{\da} \right) \ , \nonumber \\
u _{\mu} & = & i \left( u^{\da} \partial _{\mu} u -  u\partial _{\mu} u^{\da} \right) \ ,
\ea
whose transformation laws read
\ba 
\Gamma_\mu & \rightarrow & \Gamma'_{\mu} = W \Gamma_\mu W^\dag + W \pa_\mu W^\dag \nonumber \\
u_\mu &\rightarrow & u'_\mu=W u_\mu W^\dag \ . 
\ea
Finally, the covariant derivative which acts on the heavy meson field reads
\be \nabla_{\mu}   = \partial_\mu - \Gamma_\mu \ . \ee

With these ingredients, the leading-order (LO) chiral
Lagrangian $\mathcal{L}^{(1)}$ involving heavy mesons and pseudoscalar Goldstone bosons is given by

 \ba
\mathcal{L}^{(1)} & = &  \nabla ^{\mu} P \, \nabla _{\mu}P^{\da} - m_B ^2 P P^{\da} -  \nabla ^{\mu} P^{\ast \nu} \,\nabla _{\mu} P^{\ast \da}_{\nu} \nonumber \\
& &  + m_{ B} ^2 P^{\ast \mu}  P^{\ast \da}_{\mu} + ig \left(  P^{\ast \mu} u_{\mu}  P^{ \da} - P u^{\mu} P^{\ast
\da}_{\mu} \right) \nonumber \\
& & + \frac{g}{2 m_B} \left(  P^{\ast}_{\mu} u_{\alpha}   \nabla_{\beta} P^{\ast \da}_{\nu} -  \nabla_{\beta} P^{\ast}_{\mu} u_{\alpha}  P^{\ast \da}_{\nu} \right) \varepsilon ^{\mu \nu \alpha \beta} \ ,  \nonumber \\
\label{lag1}
\ea
where $P=(B^{-},\bar{B}^{0}, \bar{B}^{0}_{s})$ and  $P^{\ast} _{\mu}=(B^{*-},\bar{B}^{*0}, \bar{B}^{*0}_{s})_{\mu}$ are the SU(3) antitriplets of spin-zero and spin-one $B$ mesons with the chiral limit masses $m_B$ and $m_{B^*}$, respectively.
This Lagrangian is indeed invariant under $SU(3)_L \times SU(3)_R$ symmetry.
The NLO chiral Lagrangian $\mathcal{L}^{(2)}$ reads
\ba
\mathcal{L}^{(2)} & = & - h_0  P P^{\da} \langle \chi_+ \rangle 
+ h_1  P \chi_+ P^{\da} + h_2  P P^{\da} \langle u^{\mu}u_{\mu} \rangle \nonumber \\
& & + h_3  P u^{\mu}u_{\mu} P^{\da} + h_4  
\nabla _{\mu} P  \,  \nabla _{\nu} P^{\da}  \langle u^{\mu}u^{\nu} \rangle \nonumber \\
& & + h_5  \nabla _{\mu} P   \{ u^{\mu}, u^{\nu} \}
\nabla _{\nu}P^{\da} + \tilde{h}_0  P^{\ast \mu} 
P^{\ast \da}_{\mu} \langle \chi_+ \rangle \nonumber \\
& & - \tilde{h}_1  P^{\ast \mu} \chi_+ 
P^{\ast \da}_{\mu}  - \tilde{h}_2  P^{\ast \mu}  
P^{\ast \da}_{\mu}  \langle u^{\nu} u_{\nu} \rangle \nonumber \\
& &  - \tilde{h}_3  P^{\ast \mu}  u^{\nu}u_{\nu} 
P^{\ast \da}_{\mu} - \tilde{h}_4  \nabla _{\mu}
P ^{\ast \alpha}  \,  \nabla _{\nu}P^{\ast \da}_{\alpha}
  \langle u^{\mu}u^{\nu} \rangle \nonumber \\
& & - \tilde{h}_5 \nabla _{\mu}P ^{\ast \alpha} 
 \{ u^{\mu}, u^{\nu} \}\nabla _{\nu}P^{\ast \da} _{\alpha} \ ,
\label{lag2}
\ea
where
\be
\chi _{+} =  u^{\da} \chi u^{\da} +u \chi u \ 
\label{u_chi}
\ee
and $\chi = \mathrm{diag}(m^2_{\pi}, m^2_{\pi}, 2 m^2_{K} -m^2_{\pi})$ being the Goldstone boson mass matrix. The twelve parameters $h_i, \tilde{h}_i (i=0,...,5)$ are the low-energy constants (LECs), to be determined. However, we can make use of some constraints to reduce the set of free LECs. First, it should be noticed that in the limit of large number of colors ($N_c$) of QCD~\cite{'tHooft:1973jz}, single-flavor trace interactions are dominant. So, we fix $h_0 = h_2= h_4 = \tilde{h}_0 = \tilde{h}_2 = \tilde{h}_4 = 0$ henceforth. Besides, by imposing the heavy-quark symmetry, it follows that $\tilde{h}_i \simeq  h_i $.

In the following, we use the lowest order of the perturbative expansion of the
quantities $\Gamma _{\mu}$, $u _{\mu}$ and $\chi _{+} $ in Eqs. (\ref{lag1}) and
(\ref{lag2}), and construct the scattering matrix of the interaction between
the bottomed mesons and the pseudoscalar Goldstone bosons.

\section{\label{sec:scatt_matrix}Scattering matrix for the bottomed meson
in the meson gas} 

\subsection{Scattering matrix elements}

\begin{figure}[th]
\centering
\includegraphics[width=8.5cm]{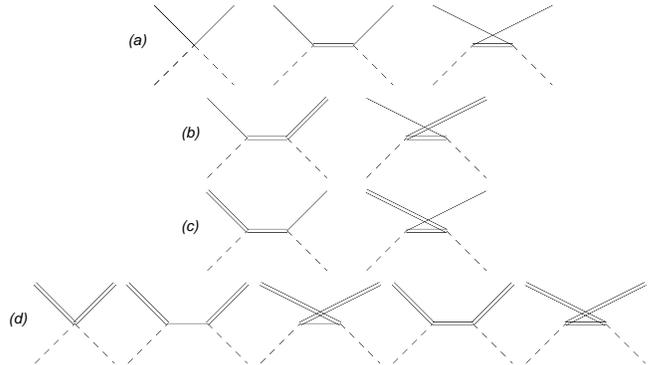}
\caption{Tree-level diagrams relevant to the scattering amplitudes for $B^{(\ast)} \phi \rightarrow B^{(\ast)} \phi $ processes. The solid, double and dashed lines represent the $B$ mesons, $B^{\ast}$-mesons and Goldstone bosons, respectively.}
\label{tree}
\end{figure}

With the Lagrangian in Eqs.~(\ref{lag1}) and (\ref{lag2}) we are able to calculate the scattering
of pseudoscalar Goldstone bosons $\phi$ off pseudoscalar $B$ mesons as well as
vector $ B^{\ast}$ mesons. In Fig. \ref{tree} we show the tree-level
diagrams constructed from the LO and NLO interactions. These include both
contact interactions and Born exchanges. 

Note that the Lagrangian density in Eq.~(\ref{lag2}) has been manifestly
constructed maintaining chiral symmetry, that is then broken only
carefully in perturbation theory upon expanding in fields and derivatives to
construct the LO and NLO chiral amplitudes. However, since the bottomed mesons
are heavy fields, the heavy-meson symmetry should be recovered by taking $m_B\to
\infty$. We discuss below the implications of this limit in the scattering
amplitude.

The spin-changing amplitudes $B^*\phi \to B\phi$ and $B\phi\to B^*\phi$, shown
in diagrams (b) and (c) of Fig.~\ref{tree}, should vanish in the limit $m_B\to
\infty$, since a collision with a Goldstone boson cannot change the heavy-quark
spin, that decouples in this limit. It can be easily proved that indeed these
amplitudes are subleading in $1/m_b$~\footnote{For a detailed discussion of
similar processes involving charmed mesons, see Ref. \cite{LADCFLEJT:2011}.}.

Turning to the elastic $B\phi$ and $B^*\phi$ amplitudes, displayed in diagrams
(a) and (d) of Fig.~\ref{tree}, one finds \cite{LADCFLEJT:2011} that the Born
exchange terms (proportional to $g^2$ with an intermediate bottomed meson
propagator) are subleading in HQET, and therefore are suppressed by $m_B^{-1}$
with respect to the contact interaction.

Hence, the final amplitude for scattering off a bottom quark in the light meson
gas, at NLO in the chiral expansion and LO in the heavy quark expansion,
irrespective of whether the heavy quark is in a $B$ or a $B^*$ meson state, is given
by
\ba
\label{ampl} V &\simeq&   \frac{C_{0}}{2 F^2} (s - u)  + \frac{2C_{1}}{ F^2} h_1    +  \frac{ 2 C_{2}}{ F^2} h_3 (p_2 \cdot p_4 ) \\
\nonumber & + &  \frac{2 C_{3}}{ F^2} h_5 \left[ (p_1 \cdot p_2 ) (p_3 \cdot p_4 ) + (p_1 \cdot
  p_4 )(p_2 \cdot p_3 ) \right] \ ,
\ea
where $ C_{i}\;(i=0,...,3)$ are channel-dependent numerical coefficients in
isospin basis, as collected in Table~\ref{table2} (channels are denoted as
$B^{(\ast)} \phi \,(I)$, with total isospin $I$). 
\begin{table}[h]
\begin{tabular}{|c ||c |c |c |c |c |c |c|}
\hline 
$C_{i}$ &  $B \pi(\frac{1}{2})$ & $B\pi(\frac{3}{2})$ & $B K(0)$ & $B K(1)$
 & $B \bar{K}(0)$ & $B \bar{K}(1)$ & $B\eta (\frac{1}{2})$ \\
\hline
$C_0 $  & -2 & 1  & -1  & 1  &  -2  & 0 & 0      \\
$C_1 $  & $ - m_{\pi} ^2$ & $ - m_{\pi} ^2$ & $  m_{K} ^2$ & $ - m_{K} ^2$ 
& $ - 2 m_{K} ^2$ & 0 & $ - m^2_\pi/3$  \\
$C_2 $  & 1 & 1 & -1 & 1 & 2 & 0 & 1/3        \\
$C_3 $  & 1 & 1 & -1 & 1 & 2 & 0 & 1/3       \\
\hline
\end{tabular}
\caption{ \label{table2}
Coefficients of the scattering amplitudes for the $B^{(\ast)} \phi \,(I)$ channels with total isospin $I$ in Eq. (\ref{ampl}). }
\end{table}

\subsection{Unitarized scattering amplitude}

It is well known that ChPT, involving a perturbative expansion up to a certain
order, is bound to work properly only at low energies. Particularly, it cannot
describe the presence of resonances in specific scattering channels, since a
resonance shows up as a pole in the $S$-matrix complex energy plane. Moreover,
the use of ChPT amplitudes at a finite order leads to cross sections which increase
monotonically with energy, eventually violating Froissart bounds imposed by
the unitarity of the $S$ matrix. This, in
turn, limits the applicability of the theory in finite-temperature calculations,
since the higher the temperature means the more energy available for the
two-body interaction. 

Imposing unitarity of the scattering amplitudes solves these problems and
furthermore accounts for resonances dynamically generated from the LO
amplitudes in those channels where the interaction is attractive.
In the present case, and motivated by our former experience in the charm sector
\cite{LADCFLEJT:2011}, we can expect to find resonances in the $B^{(\ast)} \pi
\,(1/2)$ and the $B^{(\ast)} K \,(0)$, $B^{(\ast)} \bar K \,(0)$ $S$-wave channels,
cf.~Table~\ref{table2} (we briefly review the experimental situation of the
$B$-meson excitation spectrum in Sec.~\ref{sec:lecs}). As a resonant interaction implies a more efficient
diffusion (and, thus, shorter thermalization times), we believe unitarization of
the scattering amplitudes is mandatory in this approach.

Following Refs.~\cite{LADCFLEJT:2011,Roca:2005nm}, using on-shell unitarization
via the Bethe-Salpeter equation, the
unitarized scalar amplitude $T$ can be written as
\be 
T (s) = \frac{-V (s)}{1- V (s) \ G (s)}\ , 
\label{Unitarizedampl}
\ee
where $V (s)$ is the $S$-wave projection of the scattering amplitude in Eq.
(\ref{ampl}), and $G (s)$ stands for the two-meson loop integral,
\be 
G (s) = i \int \frac{d^4 q}{(2\pi)^4} \frac{1}{(P-q)^2- m_B^{2}+i\epsilon}\frac{1}{q^2 - m_{\phi}^2 + i\epsilon} \ ,
\ee
where $m_{\phi}$ is the mass of the light meson. Employing dimensional regularization, this integral reads
\begin{eqnarray}
G (s) & = & 
\frac{1}{16 \pi^2} \left\{ a(\mu) + \ln \frac{m_B^2}{\mu^2} + \frac{m_\phi^2-m_B^2 + s}{2s} \ln \frac{m_\phi^2}{m_B^2} \right.\nonumber\\ 
& &  + \frac{q}{\sqrt{s}} \left[
\ln(s-(m_B^2-m_\phi^2)+2 q\sqrt{s})\right. \nonumber \\ 
& & + \ln(s+(m_B^2-m_\phi^2)+2 q\sqrt{s}) \nonumber  \\ 
& & - \ln(s-(m_B^2-m_\phi^2)-2 q\sqrt{s}) \nonumber \\  
& & \left. \left. 
-\ln(s+(m_B^2-m_\phi^2)-2 q\sqrt{s}) -2\pi i \right] 
\right\} \ , \label{propdr} 
\end{eqnarray}
where $\mu$ is the regularization energy scale, $a(\mu)$ is a subtraction
constant which absorbs the scale dependence of the integral, and $q$ is the modulus
of the light meson's three-momentum in the CM frame,
\be 
q = \frac{1}{2\sqrt{s}} \sqrt{\left[ s - (m_{\phi} + m_B )^2 \right] \left[s - (m_{\phi} -m_B)^2 \right]} .
\ee

It is worth mentioning that heavy-meson spin symmetry guarantees the same
scattering cross section for both $B\phi$ or $B^* \phi$ channel, and no further
spin averaging is needed. Then, we use in the collision rate defined in Eq.
(\ref{probdist}) the scattering matrix element given by

\be 
\sum | \mathcal{M}_{B \phi } (s,t,\chi)|^2 = |\ov{T}_{B \phi}|^2 \ ,
\ee    
\noindent
where $|\ov{T}_{B \phi}|^2$ is the isospin averaged unitarized amplitude,
namely:
\be 
|\ov{T}_{B \phi}|^2 = \frac{1}{\sum_{I} (2I+1)}\sum_{I}  (2I+1) |T^{I}|^2 \ ,
\ee
with $T^{I}$ being derived from Eq. (\ref{Unitarizedampl}) in the total isospin
basis.

For the sake of comparison with other systems of interest, we also evaluate the
corresponding cross sections in the CM frame,

\be
\sigma_{B \phi }(s) = \frac{1}{16\pi s}\ar {\mathcal{M}_{B \phi}}\ar^2 \ . \ee

\subsection{Free constants\label{sec:lecs}}
The only remaining task before proceeding with the calculation of the drag and
diffusion coefficients is
to determine the free constants of the theory from available data.

At the level of precision that we are working, the pion decay constant in the chiral limit
can be approximated by the physical value, $F=92$~MeV. The values of the meson physical
masses that we use are: $m_B = 5279$~MeV, $m_{B^{\ast}} = 5325$~MeV,
$m_{B_s} = 5366$~MeV, $m_{B_s^{\ast}} = 5415$~MeV, $m_{\pi} = 138$~MeV,
$m_K=496$~MeV and $m_{\eta}=548$~MeV \cite{Beringer:1900zz}.

\begin{figure}[ht]
\centering
\includegraphics[{height=6.0cm,width=8.5cm}]{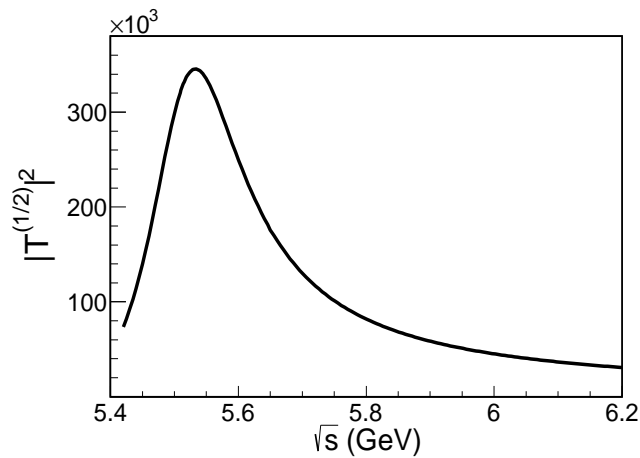}\\
\caption{\label{AmplBLO} 
Square amplitude for $B\pi$ scattering with isospin $I=1/2$ and at LO in ChPT.}
\end{figure}

Let us focus on the scattering of pions off $B$ mesons. We use the
renormalization scale $\mu=1$~GeV, and the scheme is such that the subtraction
constant $a(\mu)=-3.47$. Details in the choice of these numbers are given in
Appendix~\ref{sec:subtraction}. For this choice of parameters we show in
Fig.~\ref{AmplBLO} the squared amplitude, $|T|^2$, for $B\pi$ scattering with
isospin $I=1/2$, keeping only the LO $(s-u)$ term in the elastic amplitude $V$.
This channel, as expected, exhibits a resonant behavior with
a peak around $\sqrt{s} \simeq 5530$ MeV, which is in good agreement
with the former determinations in
\cite{Kolomeitsev:2003ac,Guo:2006fu,Flynn:2007ki}.

Notice that we have only considered the isospin channel
$I=1/2$ in the square amplitude displayed in Fig.~\ref{AmplBLO}. The exotic
$I=3/2$ channel is nonresonant in this case and in all situations presented
below.

\begin{figure}[ht]
\centering
\includegraphics[{height=6.0cm,width=8.5cm}]{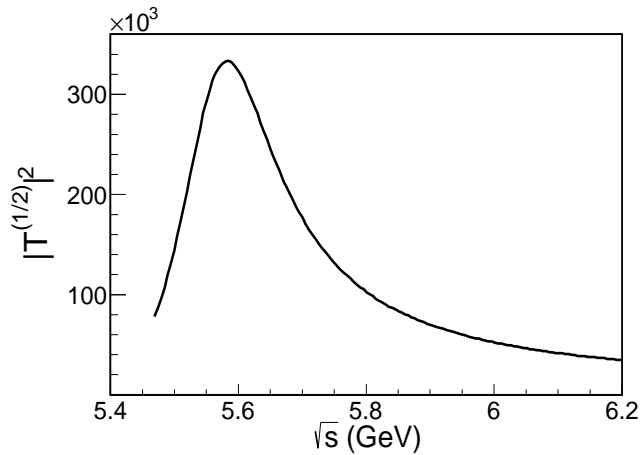}\\
\caption{\label{AmplBStarLO} 
Square amplitude for $B^*\pi$ scattering with isospin $I=1/2$ at LO in ChPT, computed replacing $m_B$ by $m_B^*$ in Eq.~\ref{Unitarizedampl}.}
\end{figure}

In addition, we compute $|T|^2$ for $B^{\ast}\pi$ scattering with $I=1/2$, just
by replacing $m_B$ by $m_B^*$ in  $V(B\pi \rightarrow B\pi)$. The result is
plotted in Fig.~\ref{AmplBStarLO}, where now the peak shows up at $\sqrt{s}
\simeq 5580$ MeV, also in agreement with Ref.~\cite{Guo:2006rp}.

Some comments about the theoretical and experimental knowledge of the $B$-meson
spectrum are in order: From heavy quark considerations (confirmed by the well
known charmed meson spectrum) one expects a spin $3/2$ doublet and a spin $1/2$
doublet as first excitations.
The former is composed by the states $B_1$ and $B_2$, both of which have
been observed as the $B_1 (5721)^0$ and $B_2^* (5747)$ resonances.
These states are narrow,
the $B_1$ state decaying mostly into $B^* \pi$ in $D$-wave.
Therefore, this state does not correspond with the resonance that is
generated in our amplitude. On the other hand, other two states $B_0$ and
$B_1$ are supposed to exist (in analogy with the $D_0$ and $D_1$ states for
charm) that 
should be much broader and have not been discovered yet. The latter decays into
$B^* \pi$ in $S$-wave, which would naturally correspond to the state that we
find in Fig.~\ref{AmplBStarLO}.
As its mass is still not determined experimentally, we can only guess that it
is very close to the
mass of the other $B_1$ state, appearing at 5720~MeV.
We justify this guess based on the fact that, in the charm sector, the masses
of the two $D_1$ states are nearly the same, and thus we expect the same
behavior for the $B$ system provided that heavy quark symmetry works. Many quarks models also predict a similar mass for the two $B_1$. See Table~\ref{tab:spectrum} for a short review.
\begin{widetext}
\begin{center}
 \begin{table}[ht]
\begin{tabular}{|c ||c |c|c||c|c|c|}
\hline 
Heavy Spin,  $J^\pi$  & D (quark model) & D (experimental) & (M,$\Gamma$) MeV & B (quark model) & B (experimental) &(M,$\Gamma$) MeV \\
\hline
1/2, $0^+$ & $D_0$ & $D_0^* (2400)$ & 2318, 267 & $\mathbf{B_0}$ & ? & ? \\
1/2, $1^+$ & $D_1$ & $D_1 (2430)$ & 2427, 384 & $\mathbf{B_1}$ & ? & ? \\
3/2, $1^+$ & $D_1$ & $D_1(2420)^0$ & 2421, 27 & $B_1$ & $B_1$(5721) &  5723, ? \\
3/2, $2^+$ & $D_2$ & $D^*_2 (2460)$ & 2466, 49 & $B_2$ & $B^*_2$ (5747) & 5743, 23 \\
\hline
\end{tabular}
\caption{ \label{tab:spectrum} List of heavy meson states. Left side: $D$ mesons. Right side: $B$ mesons. The two states in bold font are the ones that we obtain in our unitarization scheme. Experimental
data taken from Ref.~\cite{Beringer:1900zz}. Lack of experimental evidence is labeled with a question mark.}
\end{table}
\end{center}
\end{widetext}

The NLO terms, containing the $h_i$ constants, may be used to improve our
results concerning the position of the observed resonances in view of the former
discussion. $h_1$ is fixed by the mass
difference between the strange and nonstrange $B$ mesons, as obtained from the
chiral Lagrangian ${\cal L} ^{(2)}$. We have
\be
m_{B_s}^2 - m_B ^2 = - 4 h_1 (m_K ^2 - m_{\pi} ^2),
\ee
for the pseudoscalar $B$ mesons, and 
\be
m_{B_s^{\ast}}^2 - m_{B^{\ast}} ^2 = - 4 h_1 (m_K ^2 - m_{\pi} ^2),
\ee
for the vector $B$ mesons. Replacing into these equations the values of the
masses introduced above, we get $h_1 = - 1.020$ and $- 1.064$  for the scalar and
vector cases, respectively. We shall adopt in our computations an average value:
$h_1 = - 1.042$. We note that the value of $h_1$ does not yield any relevant
changes in the squared amplitudes, since the corresponding term in the NLO
amplitude is multiplied by a small $m_\pi^2$ constant.

The last free LECs to be estimated are $h_3$ and $h_5$. We proceed to fit them by
demanding that the $B^{\ast}\pi$ squared scattering amplitude peaks
at the mass of the $B_1(5721)^0$ resonance
$[I(J^P)=1/2(1^+)]; \;(5723.5 \pm 2.0)$ MeV~\cite{Beringer:1900zz}.
Following the discussion in~\cite{Guo:2009ct} we can estimate a valid range for
$h_5$ from our results in the $D$ sector.
Taking into account that $h_5$ scales as $h_5 \sim m_D^{-2}$, we use for
the $B$ sector:
\be h_5^B = h_5^D \left( \frac{m_D}{m_B} \right)^2 \ ,
\ee
leading to a value of
$h_5=-0.04$~GeV$^{-2}$. The value of $h_3$ does not affect very
much the pole position of the resonance.

\begin{figure}
\centering
\includegraphics[{height=6.0cm,width=8.5cm}]{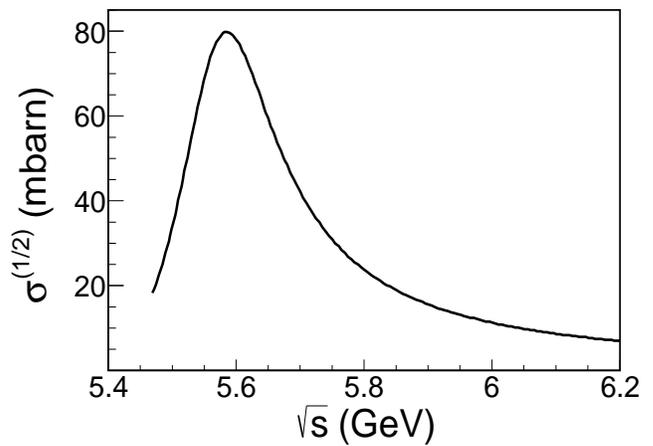}\\
\caption{\label{AmplBStarNLO} 
Cross section for $B^*\pi$ scattering with isospin $I=1/2$ and at NLO in ChPT, computed replacing $m_B$ by $m_B^*$, and considering $(h_1,h_3,h_5)=(-1.042,2.5,-0.04$ GeV$^{-2})$.
The resonance is to be understood as the broad $B_1$.}
\end{figure}

Choosing a value of $h_3=2.5$ we obtain the final result in
Fig.~\ref{AmplBStarNLO}. The predicted mass for this state from the peak of the
amplitude is still about 100~MeV below the $B_1(5721)^0$ mass.
We find, as discussed in Appendix~\ref{sec:subtraction}, that it is not possible
to fix simultaneously the low energy and subtraction constants --in a natural choice-- 
to get values nearer the expected $B_1(5721)$ mass.
Furthermore, our findings are in agreement
with those in Refs.~\cite{Guo:2006fu,Guo:2006rp,Kolomeitsev:2003ac}. 
Because of the lack of experimental evidence of a $B_1$ state decaying into $B^* \pi$, the most reasonable choice is
to keep the values of the LECs determined above.

\begin{figure}[ht]
\centering
\includegraphics[{height=6.0cm,width=8.5cm}]{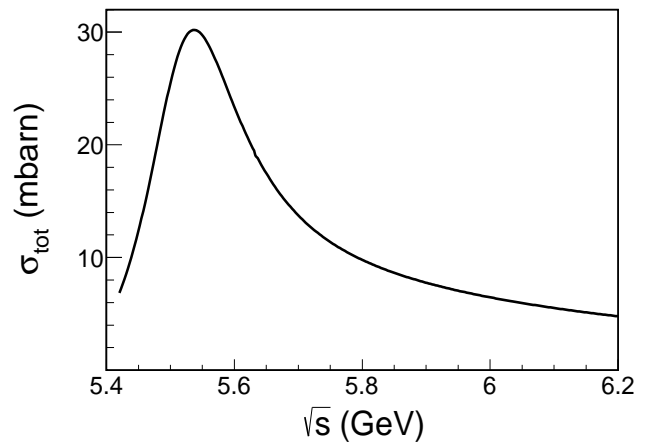}\\
\caption{\label{AmplBNLO} 
Total $B\pi$ cross section (both $1/2$ and $3/2$ isospin channels) at NLO in ChPT, $(h_1,h_3,h_5)=(-1.042,2.5,-0.04$ GeV$^{-2})$.}
\end{figure}

For completeness, we also compute $B\pi$ scattering at NLO in ChPT. This channel
is also resonant, with $m_{B_0}=5534$ MeV. Again, there are no data yet at our
disposal to compare with. The total cross section for the $B\pi$ channel
(considering the two isospin channels $I=1/2$ and $3/2$) is shown in 
Fig.~\ref{AmplBNLO}.

The $I=1/2$ cross section at threshold is 18.1~mbarn, or equivalently an
$S$-wave scattering length of $a^{1/2}_{B\pi}=0.38$~fm or $m_{\pi}
a_{B_\pi}=0.26$ is found, in total agreement with the results of \cite{Flynn:2007ki} but larger
than the result of \cite{Liu:2009uz}, $a^{1/2}_{B\pi}=0.25$~fm. For the
repulsive channel $I=3/2$ we find a cross section of 1.5~mbarn, or
$a^{3/2}_{B\pi}=-0.11$~fm, close to the results in \cite{Liu:2009uz} of
$-0.17$~fm.

After the estimation of the free relevant constants for the $B^{(*)}\phi$
scattering amplitude, we are now in position to compute the transport
coefficients, which are the subject of the next section. 

\section{\label{sec:transport}Drag and diffusion coefficients}

We proceed to calculate the $F$, $\Gamma_0$ and $\Gamma_1$ transport
coefficients defined in Eq. (\ref{Transportintegrals}), for bottomed mesons in a
pure pion gas. We expect the contribution of pions to be the most relevant one because of their large
multiplicity in comparison to other particles. For completeness, in this work we
also account for the effect of the other members of the light meson $SU(3)$
octet. A detailed discussion is done in Sec.~\ref{sec:kaons}.

\begin{figure}
\centering
\includegraphics[{height=6.0cm,width=8.5cm}]{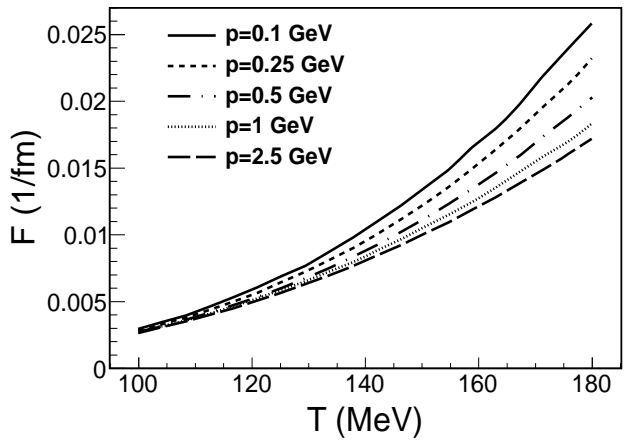}
\caption{\label{fig:dragP0} 
Momentum-space drag coefficient as function of temperature for several bottom quark momenta in the pion gas.}
\end{figure}

In Fig.~\ref{fig:dragP0} we show the dependence on the temperature and $B$-meson
momentum of the drag coefficient $F$. We observe an increase of a factor of
about 6-8 in the range from $T=100-180$~MeV, which means that the
temperature evolution of the thermal medium can modify the nature of its
interaction with the $B$ mesons.
So, the drag in a heavy-ion collision is considerably
strengthened in the hotter stages, with a significant interaction between heavy
mesons and thermal medium, and accordingly at larger momentum transfers.
However, as the temperature of thermal bath diminishes, the magnitude of
interaction decreases and the bottomed mesons move more freely.
The drag coefficient exhibits a mild momentum dependence of about 10\% in the range $[0.5,2.5]$~GeV,
whereas it is more pronounced at low $p$.
From this coefficient we estimate the relaxation length of bottom quarks in the
hadronic medium.
Taking the pion gas at a
temperature of 150 MeV (at which our approach is reliable), we find
\be
\lambda_B (T= 150 \textrm{ MeV}, p = 1 \textrm{ GeV}) = \frac{1}{F}\simeq \frac{1}{0.01}\textrm{ fm} = 100\textrm{ fm}
\label{relax}
\ee
for bottomed mesons traveling with a typical momentum of 1~GeV. This value of
$\lambda_B$ is considerably bigger than the case of charmed mesons, evaluated in
the same approach in Ref.~\cite{LADCFLEJT:2011} (there $\lambda_D\simeq 40$~fm).
Thus, it is reasonable to expect that the lifetime of the pion gas (typically
5-10~fm) is smaller than the relaxation time of both $D$ and $B$ mesons, which
means that heavy quarks do not completely relax before leaving the hadronic
medium.

Another point worthy of mention is that the drag coefficient for charmed mesons
computed in our previous work~\cite{LADCFLEJT:2011} is about 3 times larger
than the one for bottomed mesons at small $p$. This scaling is the correct one when
looking at the nonrelativistic expression of this coefficient in the static limit
\be
 \label{eq:nonrelF} 
 F \simeq \frac{1}{3} \sigma P \sqrt{\frac{m_{\pi}}{T}} \frac{1}{m_B} \ ,
\ee
where $\sigma$ is the total cross section and $P$ the pressure of the gas.
Note that there is a dependence on the heavy-flavor mass in the denominator. This
makes the drag coefficient smaller for heavier mesons. With respect to the $D$
meson system, the drag force is suppressed by a factor of $m_B/m_D \simeq 2.8$
in very good agreement with what we observe in our results. Note that in the low
energy limit the cross section does not depend on the heavy meson mass. This is
not the case at higher temperatures where Eq.~(\ref{eq:nonrelF}) contains
further corrections in powers of $1/m_B$. 
This scaling is also observed for the
drag coefficient of $b$ and $c$ quarks in the static limit beyond the critical
temperature within the phenomenological approach of Ref.~\cite{Rapp:2008qc,vanHees:2007me}, in
total accordance with our findings in the hadronic phase. We observe, however,
that this scaling is not maintained in the results of
\cite{Ghosh:2011bw,Das:2011}. There, a similar approach based on heavy-meson
ChPT amplitudes is used to calculate the transport coefficients of $D$- and
$B$ mesons. However, the authors employ NLO perturbative amplitudes in the case
of $D\pi$ scattering \cite{Ghosh:2011bw}, overestimating the effect due to the
high-energy dependence of their cross sections. This is partly solved in
\cite{Das:2011} for the $B$ system, where NLO amplitudes at threshold energy
(scattering lengths) are used in the evaluation of $B$-meson transport, thus
taming the high-energy behavior of the amplitudes. In contrast, this
approximation
underestimates diffusion from resonant scattering in the $B^{(*)}\pi (1/2)$
channel. This discussion reinforces the role of unitarization of low-energy
scattering amplitudes to obtain realistic transport coefficients at high
temperatures in the heavy-flavor sector. A more detailed discussion of the use
of perturbative versus unitarized amplitudes is carried over at the end of this
section.

\begin{figure}
\centering
\includegraphics[{height=6.0cm,width=8.5cm}]{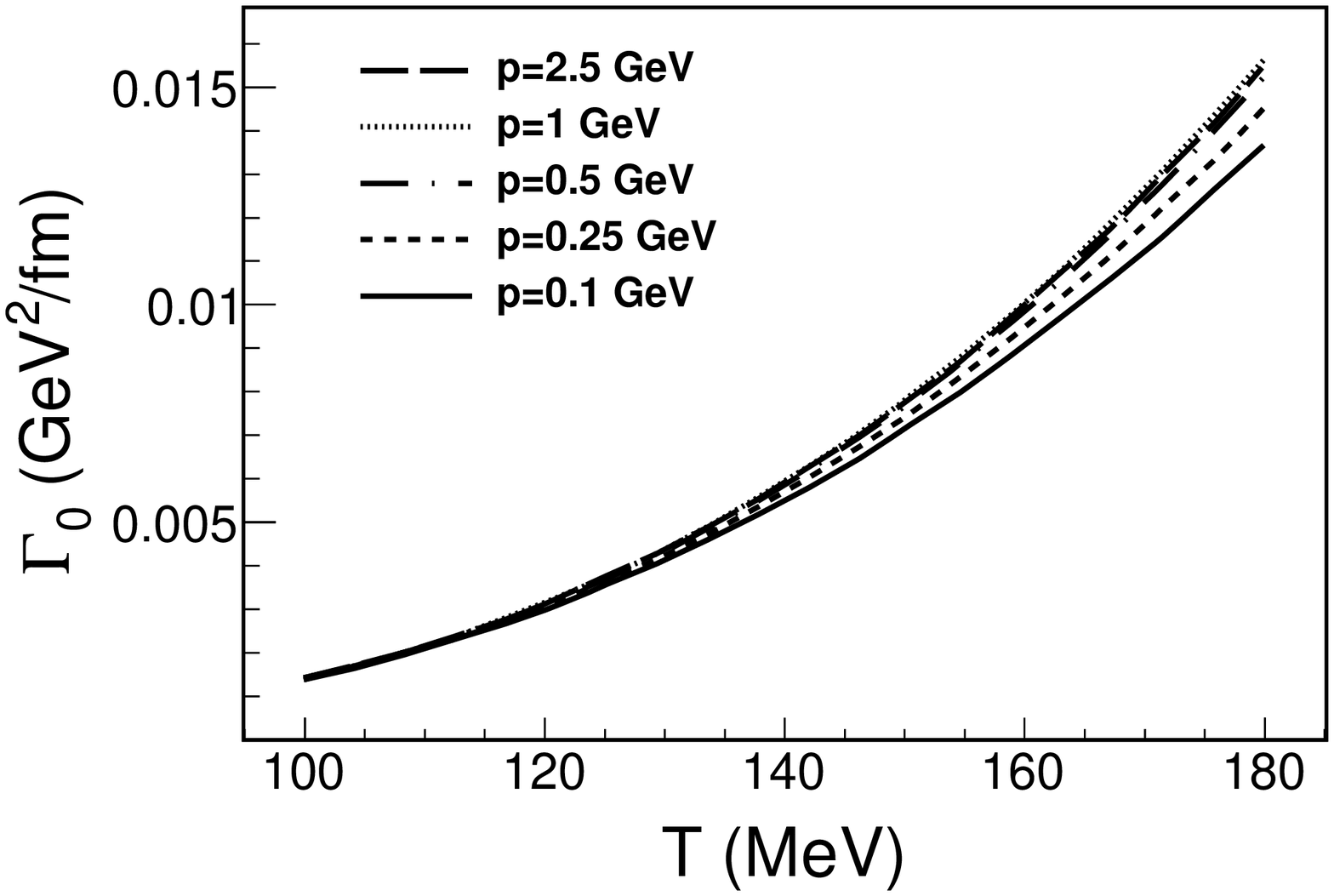}
\includegraphics[{height=6.0cm,width=8.5cm}]{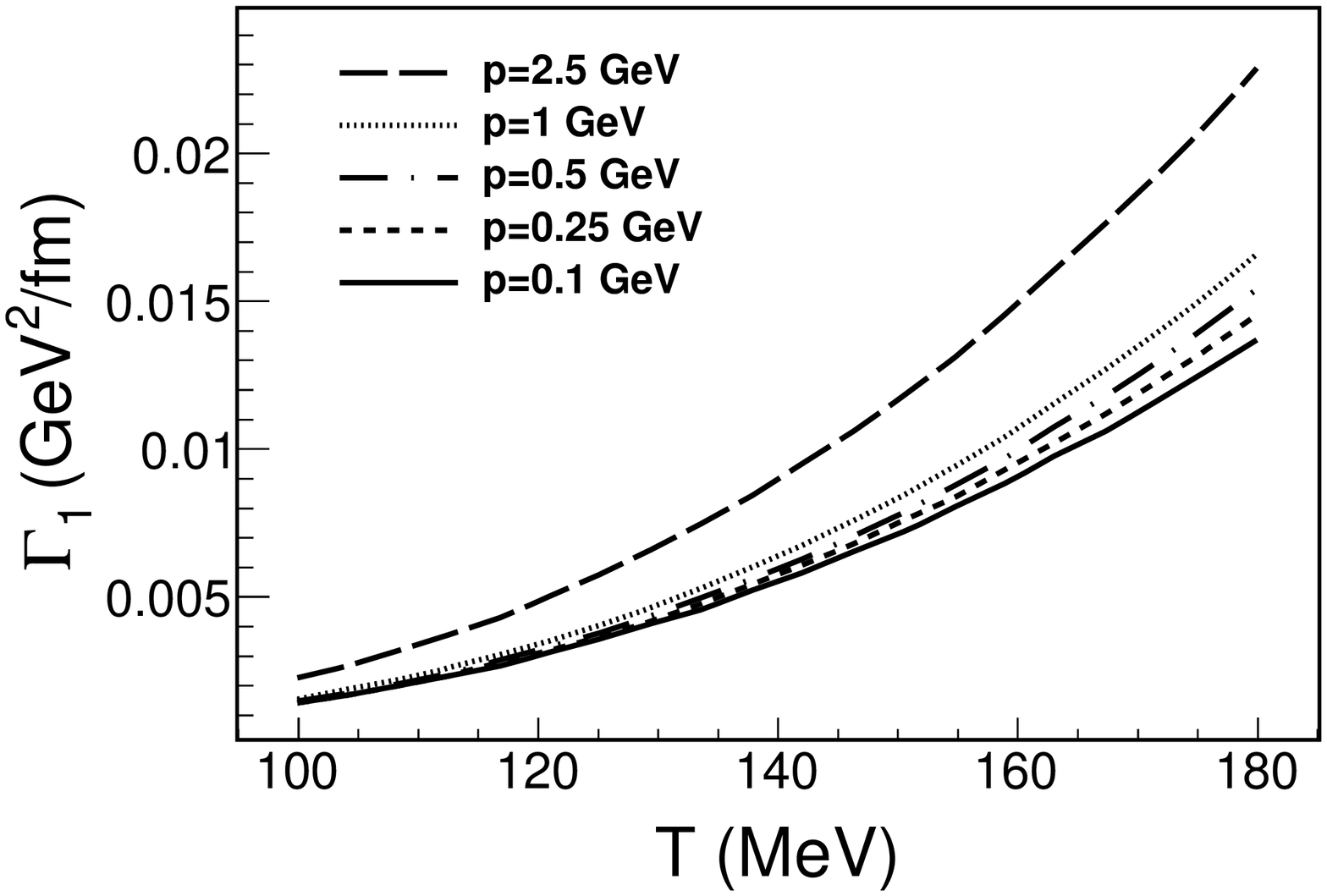}
\caption{\label{fig:GP0} 
Momentum-space diffusion coefficients as function of temperature for several
bottom quark momenta in the hadron gas.}
\end{figure}

In Fig.~\ref{fig:GP0} the temperature dependence of the $B$
meson diffusion coefficients $\Gamma_0$ and $\Gamma_1$ is displayed for several
momenta.
The two
coefficients become degenerate in the static limit, $p\rightarrow 0$. At $p=0.1$
GeV this limit is already well reached. Interestingly, one finds that
numerically this coefficients are very similar to those obtained for the $D$
meson case in Ref.~\cite{LADCFLEJT:2011}.
Looking at the nonrelativistic expression of these
coefficients gives a clue for this fact:
\be 
\Gamma_0 , \Gamma_1 \simeq \frac{1}{3} \sigma P \sqrt{m_{\pi} T} \ , 
\ee
where we have used the Einstein relation
\be 
\label{Einstein} 
F = \frac{\Gamma}{m_B T}  
\ee
to obtain this equation.
The coefficients $\Gamma_0$ and $\Gamma_1$ are independent of the heavy quark
mass in the nonrelativistic limit. For this reason the results in
Fig.~\ref{fig:GP0} are basically the same as those in
Ref.~\cite{LADCFLEJT:2011}. In Ref.~\cite{Rapp:2008qc,vanHees:2007me} very similar diffusion
coefficients are found in the $c$ and $b$ quark systems, confirming the
prediction of the nonrelativistic kinetic theory. These findings constitute an
additional consistency test of our calculations.

\begin{figure}
\centering
\includegraphics[{height=6.0cm,width=8.5cm}]{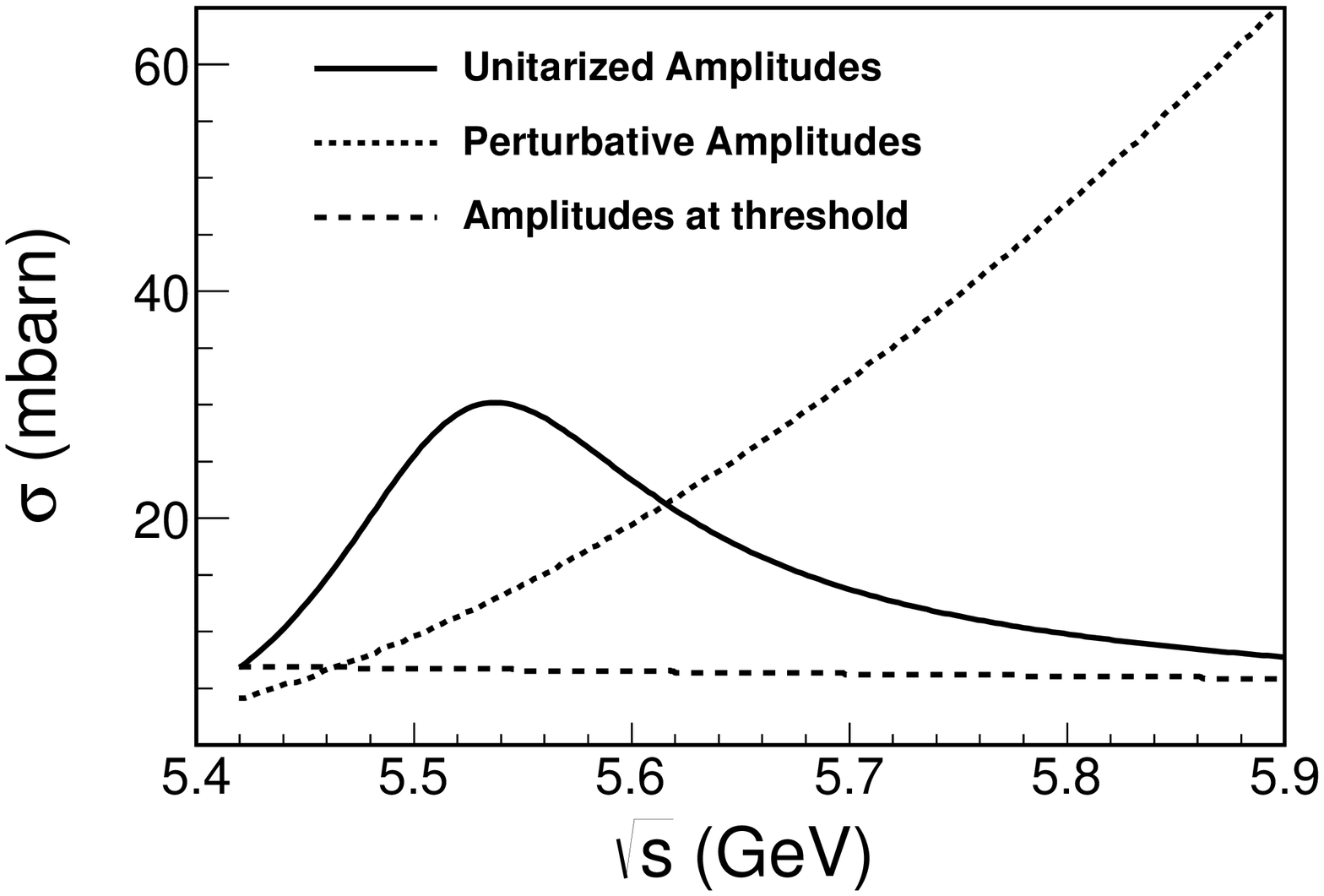}
\includegraphics[{height=6.0cm,width=8.5cm}]{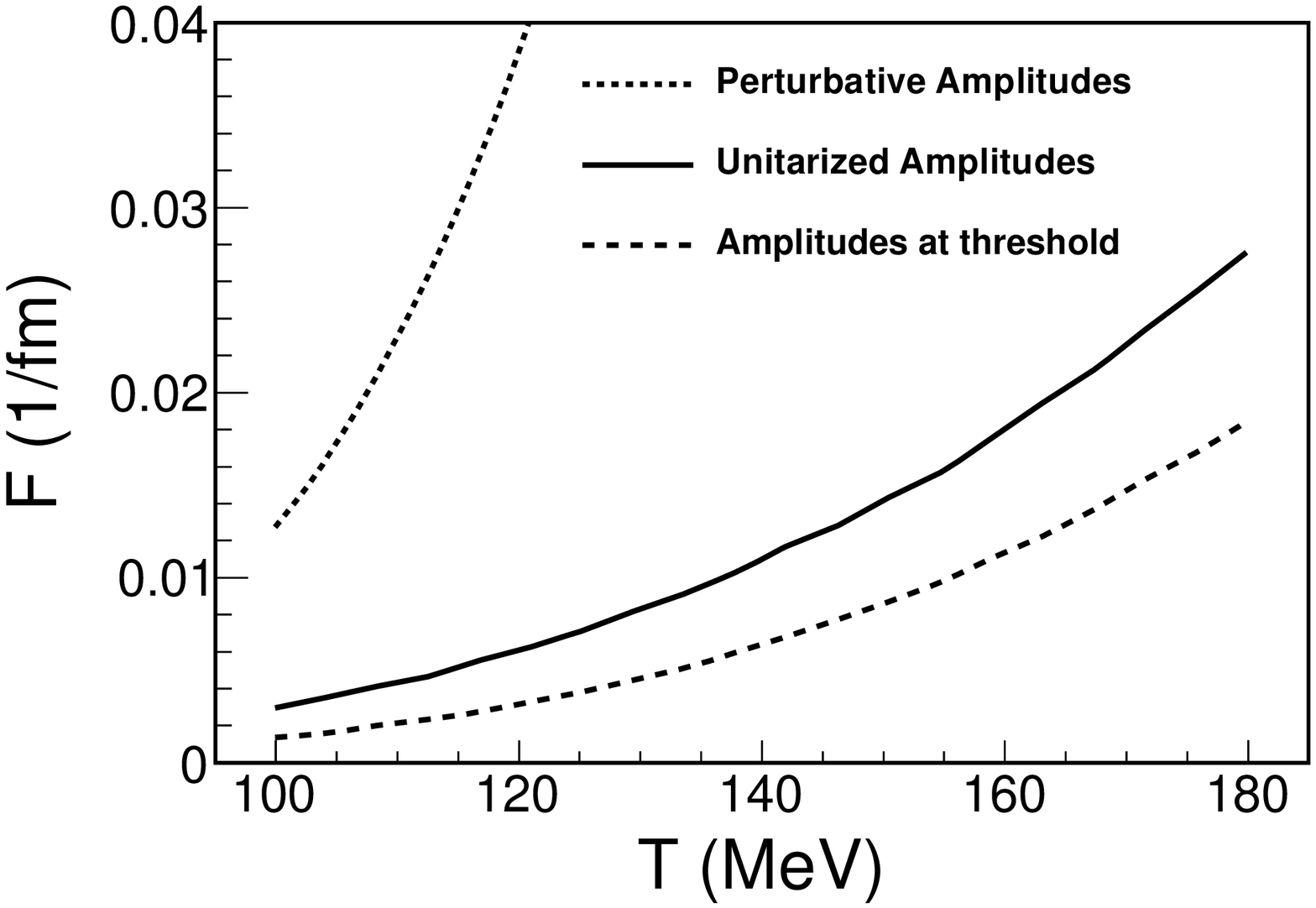}
\caption{\label{fig:unit-vs-pert} 
Upper panel: $B\pi$ isospin averaged total cross section for
different choices of the scattering amplitudes: unitarized, perturbative, and
scattering lengths. Lower panel: Corresponding drag coefficient in static limit
as a function of temperature.
}
\end{figure}

In the following we study the role of unitarization in obtaining realistic
amplitudes for $B$ meson scattering off the light gas and its impact on the
transport coefficients. We deem this is an important point in order to
understand the wide range of results for transport coefficients that can be
found in literature within similar approaches based on effective theories of
heavy meson interactions. In Fig.~\ref{fig:unit-vs-pert} we depict the $B\pi$
total cross section in three different schemes: using NLO (perturbative)
chiral amplitudes, unitarized amplitudes as described in
Sec.~\ref{sec:scatt_matrix}.B, and NLO amplitudes evaluated at threshold energy
(scattering lengths). We observe that the perturbative cross section grows
monotonically with energy, as expected from the chiral expansion. The cross
section within the threshold approximation is essentially flat over the full
energy range. The unitarized cross section, in contrast, peaks at the resonance
region, dominating over the other two schemes, whereas later it decreases at
higher energies as expected from phase space considerations.
The corresponding drag coefficient with $p=0.1$ GeV is shown in the lower panel of Fig.~\ref{fig:unit-vs-pert}.
The perturbative scheme leads to an unrealistic temperature behavior of $F$, which is clearly tied to the unphysical
high-energy behavior of the cross section. The scattering length approach does
not suffer from this artifact at high energies.
However, it provides a rather smaller diffusion
coefficient (factor $1.5-2$) over the whole temperature range as compared to the
unitarized scheme, since it misses the
$s$-channel enhancement of the interaction due to the presence of resonances.
The unitarized scheme leads to the most realistic result in our
opinion, accounting for the phenomenology of the heavy-meson interaction in view
of the current knowledge of the $B$-meson spectrum, and with a
controlled high-energy behavior.

\begin{figure}
\centering
\includegraphics[{height=6.0cm,width=8.5cm}]{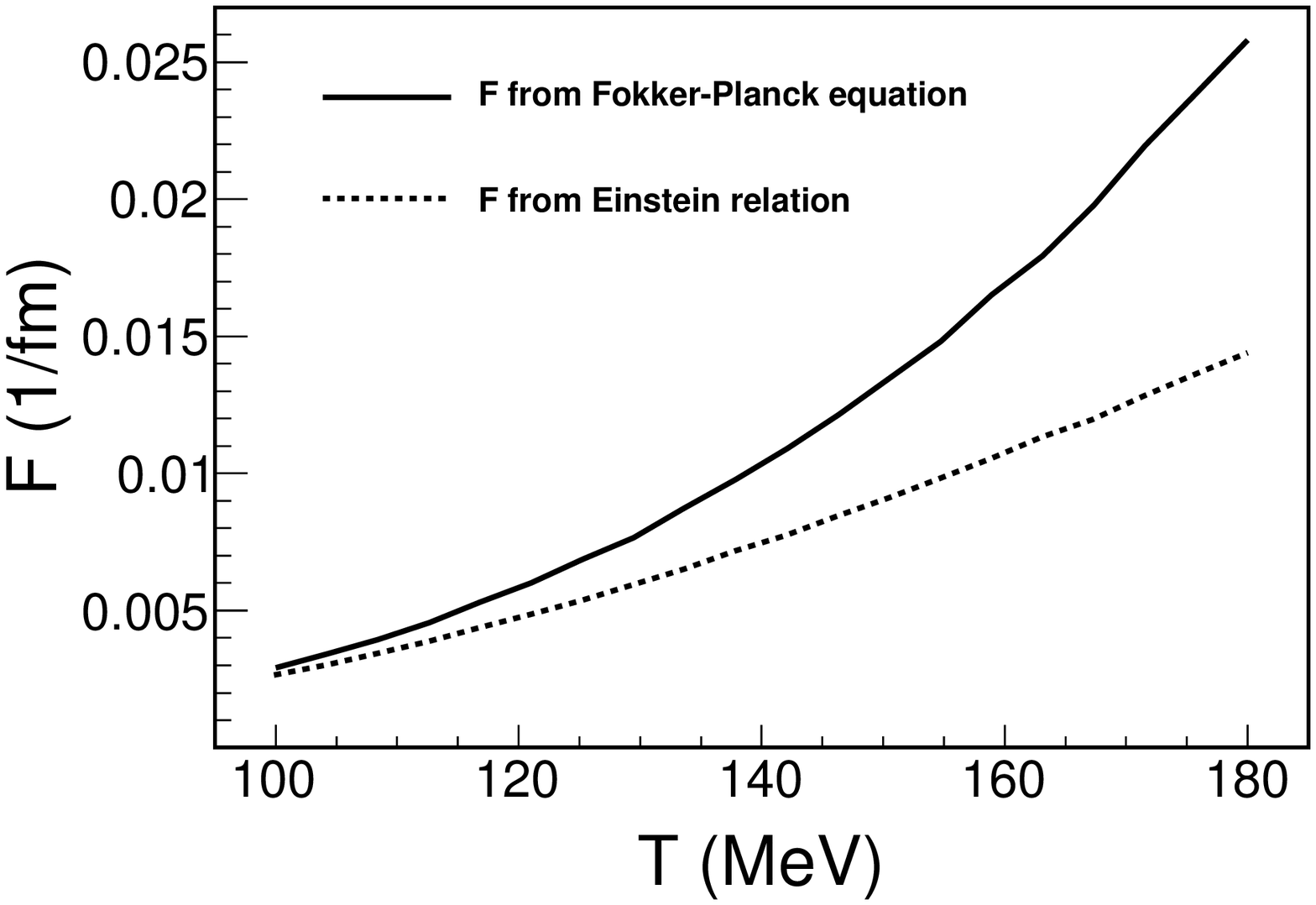}
\includegraphics[{height=6.0cm,width=8.5cm}]{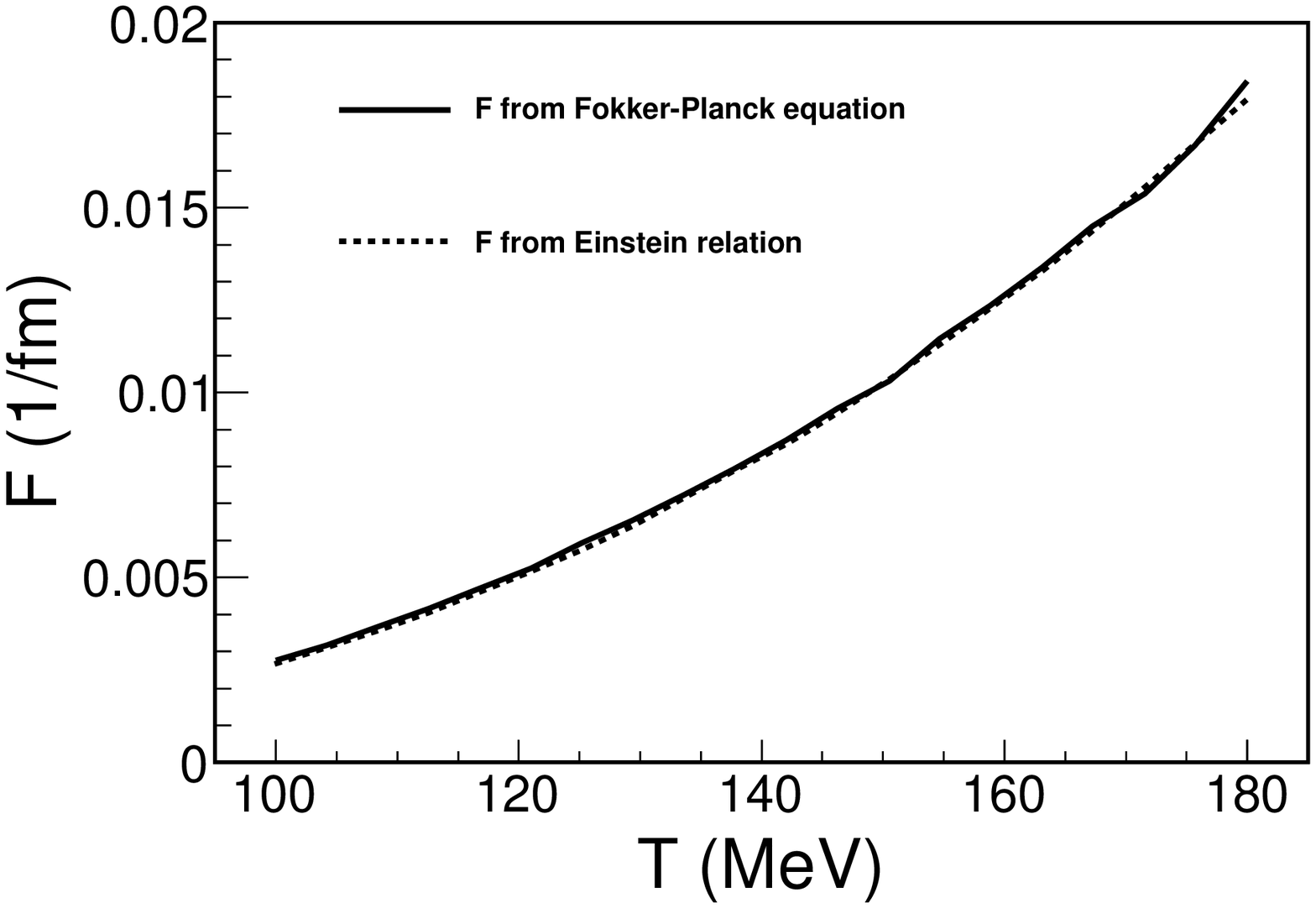}
\caption{\label{fig:F_wocut} Upper panel: Drag force at $p=0.1$ GeV as evaluated from
Eq.~(\ref{Transportintegrals}) and ~(\ref{Einstein}). The Einstein relation is
not satisfied at high temperatures due to the momentum cutoff coming
from the effective theory ($\sqrt{s}\lesssim 6.5$ GeV). Lower
panel: Same as above but with a larger cutoff momentum.
}
\end{figure}

Finally, a comment on the Einstein relation within our
calculation is in
order. In the evaluation of $F$, $\Gamma_0$ and $\Gamma_1$ we have not made use
of the Einstein relation to obtain the coefficients at low momentum, but
performed the calculation explicitly as in
Eqs.~(\ref{Transportintegrals}) and (\ref{probdist}). A pertinent consistency test can
be done by obtaining the diffusion coefficient from the drag coefficient (or
viceversa) in static limit from the Einstein relation. Numerically, the
integrations in Eq.~(\ref{Transportintegrals}) have to be performed with a
suitable pion momentum cutoff, which is the scale at which 
the effective theory ceases to be valid.
The presence of this cutoff is expected not to affect the integrations (at least at low temperatures)
because the distribution function supresses the integrand at large pion momentum 
so that high-energy contributions are negligible. 
In practice, the used cutoff $\Lambda_\chi \equiv 4 \pi f_{\pi} =1.2$~GeV makes our
results sensitive at high temperatures, where the Einstein relation is not
well fulfilled. 
Notice that one cannot arbitrarily increase this cutoff
to achieve convergence, as this implies that the scattering amplitudes have to
be evaluated at energies that escape the expected validity of the effective
theory. This would introduce an uncontrolled uncertainty in the coefficients. Therefore we prefer
to use the safe value of 1.2~GeV for the pion momentum cutoff.
Studying how much the static coefficients deviate from fulfilling the
Einstein relation as a function of the momentum cutoff can be considered as an
estimate of the systematic error in our computation of the transport
coefficients. An example of this is
provided in Fig.~\ref{fig:F_wocut}.
In the upper panel we plot the function $F$ at $p=0.1$~GeV (very close to the
static limit) as obtained from Eqs.~(\ref{Transportintegrals},\ref{probdist})
with a momentum cutoff of 1.2~GeV, together
with the determination from $\Gamma_0, \Gamma_1$ 
using the Einstein relation.
In the lower panel the same curves are shown
with a
cutoff of 3~GeV. This higher cutoff ensures convergence of the transport
integrals at all
temperatures and the Einstein relation is well satisfied.
It is worth mentioning that the scattering length scheme discussed above ensures
a much quicker convergence of the transport integrals, at a cost, however, of
implementing an unrealistic energy dependence of the heavy-light meson cross
sections and missing the phenomenological information from resonance-enhanced
diffusion.

\section{\label{sec:kaons}Effect of kaons and $\eta$ mesons}

Analogous to our previous work, so far we have studied the interaction of the
bottomed mesons with pions, as they are most abundant. However, the mesonic gas is also populated by kaons and $\eta$
mesons, with whom the heavy mesons can also interact. In
Sec.~\ref{sec:scatt_matrix} we only discussed scattering channels involving
pions, although we have actually extended the computation to the other members of
the $SU(3)$ octet. With this information it is straightforward to implement the
contribution from kaons and $\eta$'s to the transport coefficients. We also added
these states for the sake of comparison with other references, particularly
\cite{Das:2011}.

The changes in our calculation are minimal and they affect the collision rate
$w$ defined in Eq.~(\ref{probdist}), which now reads
\be w = w_{\pi} + w_{K}+ w_{\overline{K}} +w_{\eta} \ . \ee
We note that the contribution to the transport coefficients from different
species in the gas is always additive,
so we expect a moderate increase with the inclusion of these new states.

\begin{figure}
\centering
\includegraphics[{height=6.0cm,width=8.5cm}]{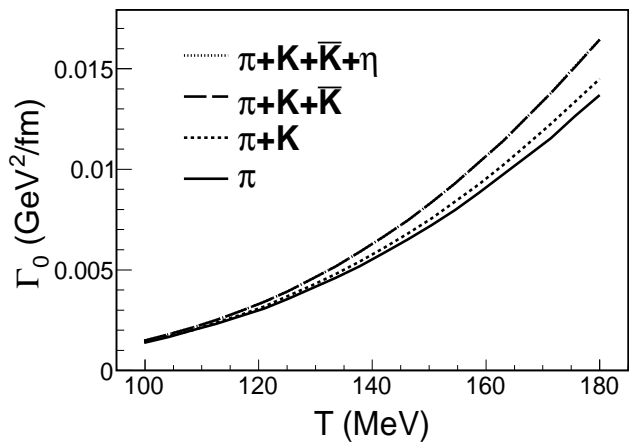}
\caption{\label{fig:G0su3} 
$\Gamma_0$ coefficient of $B$ meson with momentum $0.1$ GeV, considering the mesonic gas constituted by different particles.}
\end{figure}

In Fig.~\ref{fig:G0su3} we plot the $\Gamma_0$ coefficient of $B$ mesons with
a momentum $p=0.1$~GeV, considering the mesonic gas constituted by different
particles. We see that the most relevant contribution comes from the pion gas, as
expected. Performing a decomposition of individual hadron contributions to
$\Gamma_0$ at $T = 150$ MeV we observe that
pions provide almost 90\% of the total, while the next contribution is
provided by kaons and (mostly) antikaons. The contribution of $\eta$ mesons is 
almost negligible, in agreement with the fact that the $B \eta$
interaction vanishes at LO.

\begin{figure}
\centering
\includegraphics[{height=6.0cm,width=8.5cm}]{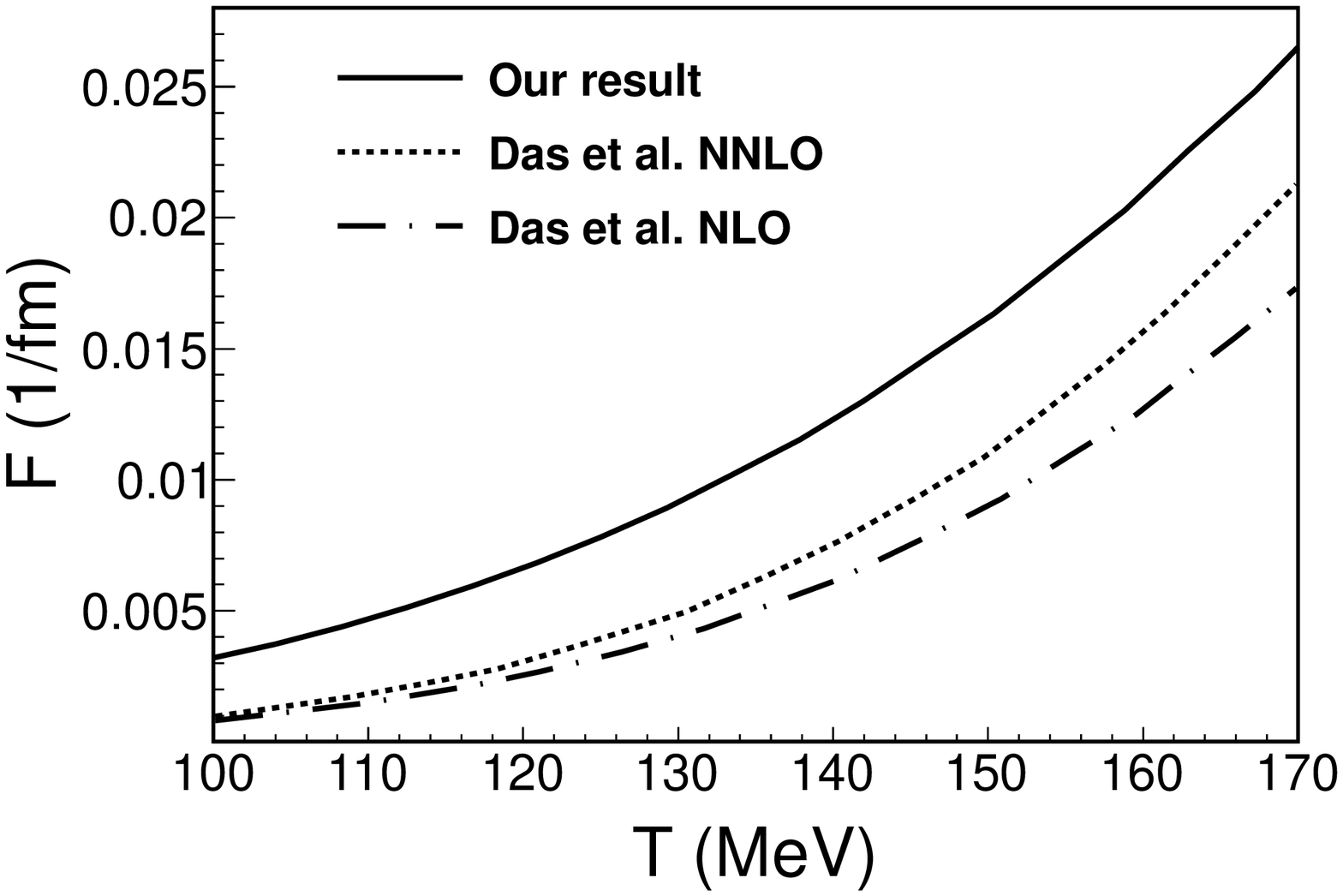}
\includegraphics[{height=6.0cm,width=8.5cm}]{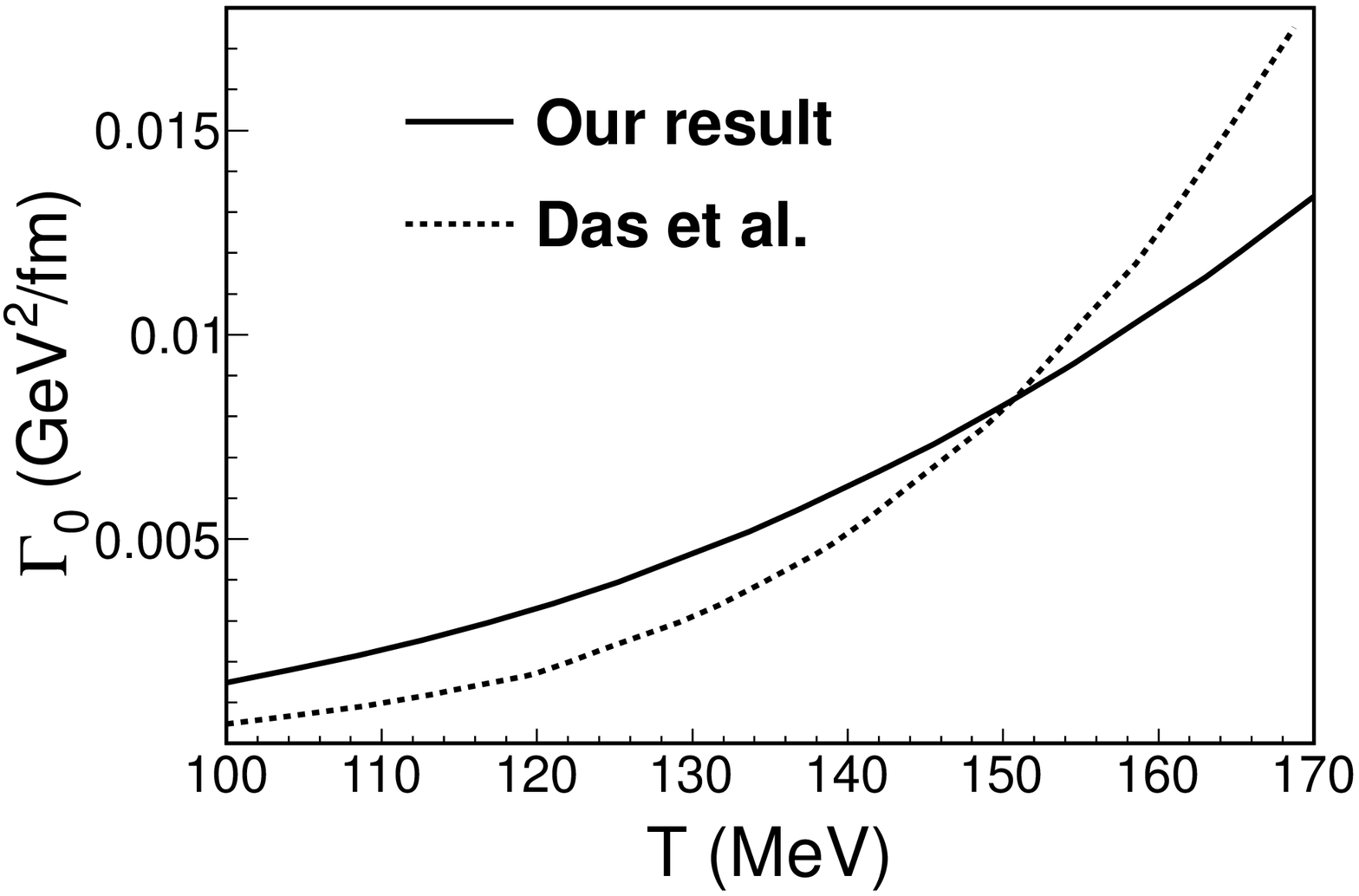}
\caption{\label{fig:FG0comp} 
$F$ and $\Gamma_0$ coefficient at $p=0.1$ GeV in comparison with the results of~\cite{Das:2011}.}
\end{figure}

In Fig.~\ref{fig:FG0comp} we compare the value of the $F$ and $\Gamma_0$
coefficients at $p=0.1$~GeV with those of Ref.~\cite{Das:2011}. It can be
noticed that our results are larger than those of \cite{Das:2011}, where no
unitarization is performed. $\Gamma_0$, though, exhibits a smoother growth with
temperature in our case, actually being overtaken beyond $T\simeq 150$~MeV. We
believe that the reason for this discrepancy is the use, in \cite{Das:2011}, of
scattering lengths (valid in principle for very low energies) in the whole
energy and temperature range. As discussed above, this will certainly bring an
underestimation of the cross section in the region where resonances take over,
producing lower transport coefficients. In contrast, at very high energies, the
use of a constant cross section may lead to the opposite effect, explaining the
behavior of $\Gamma_0$ in the high-temperature region from Ref.~\cite{Das:2011}.

Another point deserving attention is the relaxation of bottom quarks in the
present situation, for sake of comparison with that obtained in
Section~\ref{sec:transport}, which has been calculated for a pure pion gas.
Taking the meson gas at a temperature of 150~MeV, the relaxation length of
bottomed mesons traveling with 1~GeV momentum in the full meson gas is

\be
\lambda_B (T= 150 \textrm{ MeV}, p = 1 \textrm{ GeV}) = \frac{1}{F}\simeq \frac{1}{0.0124}\textrm{ fm}\simeq 81\textrm{ fm} \ , 
\label{relax2}
\ee
We find that the value for $\lambda_B $ in Eq.~(\ref{relax2}) is reduced with respect to that estimated in
Eq.~(\ref{relax}), since in the present case the inclusion of other contributions
to the meson gas yields an increase of the transport coefficients. However,
this thermal relaxation time continues greater than the lifetime of hadron gas,
allowing us to consider that indeed heavy quarks are carriers of information of
the phase transition upon exiting the hadron gas.

\begin{figure}
\centering
\includegraphics[{height=6.0cm,width=8.5cm}]{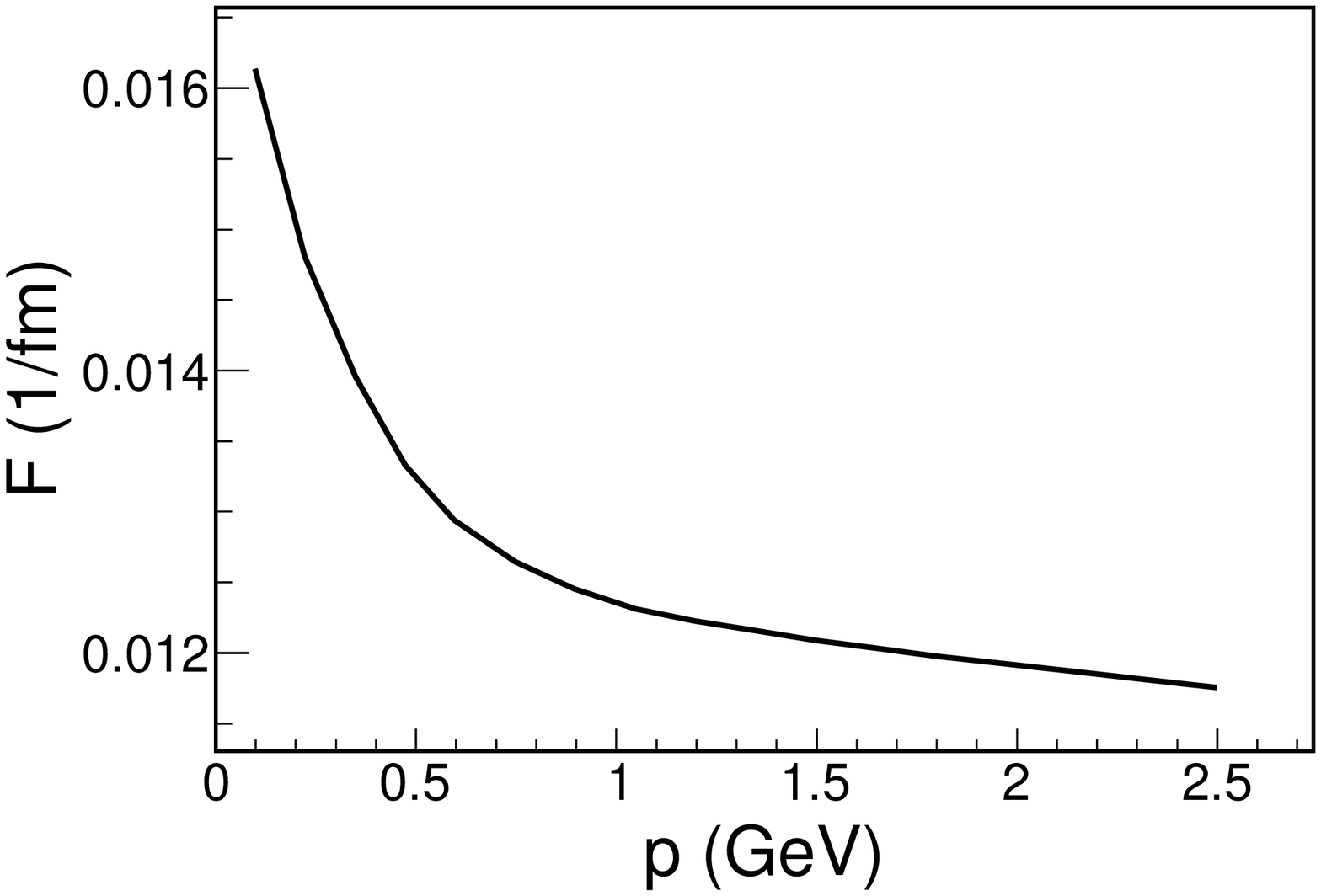}
\includegraphics[{height=6.0cm,width=8.5cm}]{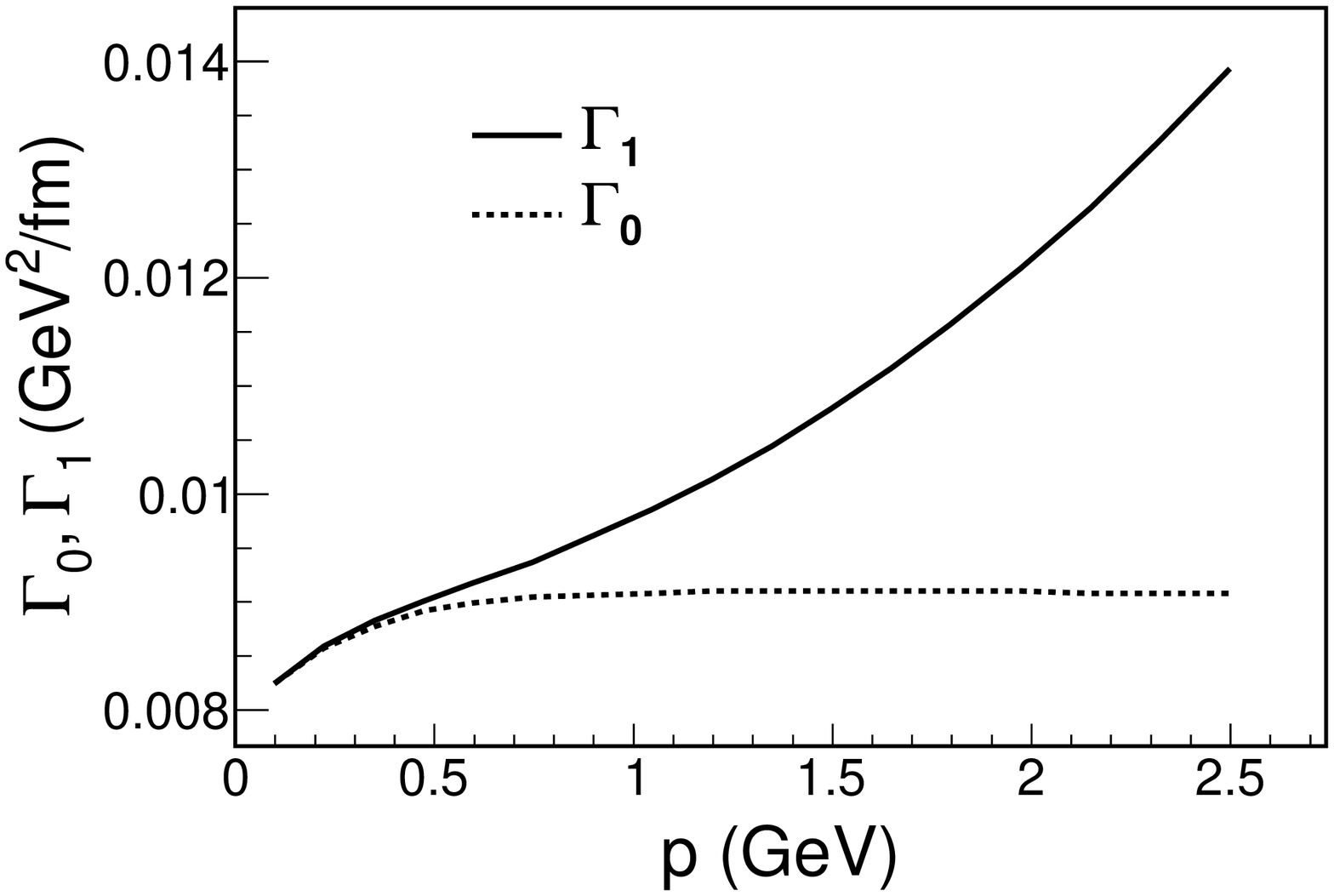}
\caption{\label{fig:FG0G1-full-gas-vs-p} 
$F$ (upper panel) and $\Gamma_0$, $\Gamma_1$ (lower panel) coefficients at
$T=150$~MeV as a function of the heavy-meson momentum.}
\end{figure}

\begin{figure}
\centering
\includegraphics[{height=6.0cm,width=8.5cm}]{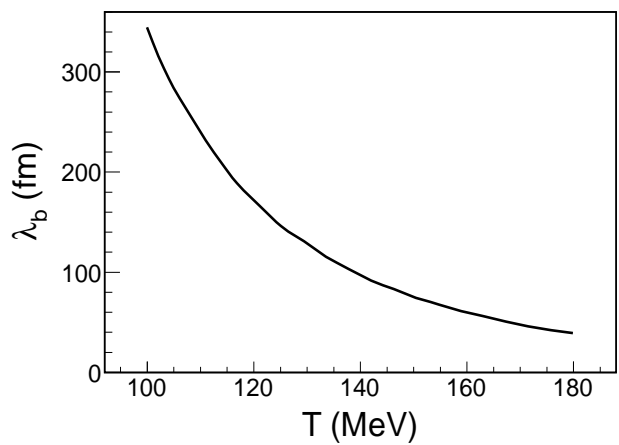}
\caption{\label{fig:lambda-full-gas-vs-p} 
Bottom relaxation length in the hadronic phase at $p=0.1$~GeV as a function of the temperature.}
\end{figure}

For completeness, we also account in Figs.~\ref{fig:FG0G1-full-gas-vs-p} and
\ref{fig:lambda-full-gas-vs-p} for the
evolution of the transport coefficients and the extracted bottom relaxation
length with the heavy-meson momentum, in the
full meson gas, at a temperature of 150~MeV. 

Notice that we have presented results in a wide temperature
range. When $T$ is of the order of the pion mass
some corrections in our scheme are in order because of (i) the loss of the
diluteness assumption in the kinetic equation, and (ii) missing medium effects in the
scattering amplitudes.
To have
an idea of the error due to this fact we have studied 
the effect of a change in the pion mass
for the case $p=1$ GeV and $T=150$ MeV.
With a 10 \% of the pion mass variation,
i.e., $m_{\pi} = 138\pm 14$ MeV, we obtain $F=1.24^{+0.03}_{-0.04}
\cdot 10^{-2}$ fm$^{-1}$, $\Gamma_0=(9.07 \pm 0.29) \cdot 10^{-3}$ GeV$^2$/fm and
$\Gamma_1=9.75^{+0.29}_{-0.31} \cdot 10^{-3}$ GeV$^2$/fm, which represent a
variation of 3.2 \% around the central value in the three cases.
We thus expect large temperature corrections to be relatively small. In view of
the latter, we have assigned
an uncertainty to the relaxation length in Eq.~(\ref{relax2}):
$\lambda_B=81 \pm 2$ fm.

\begin{figure}
\centering
\includegraphics[{height=6.0cm,width=8.5cm}]{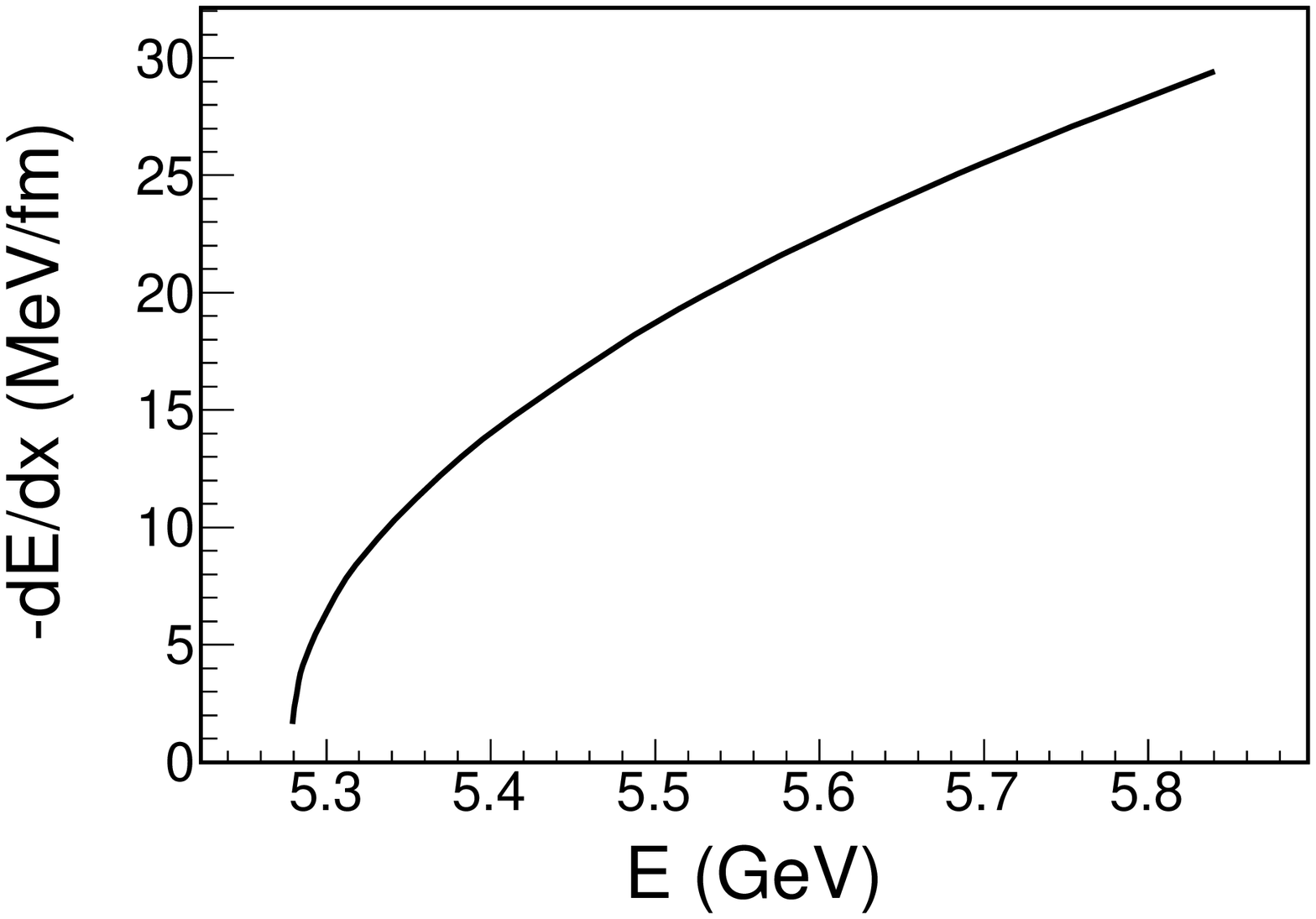}
\includegraphics[{height=6.0cm,width=8.5cm}]{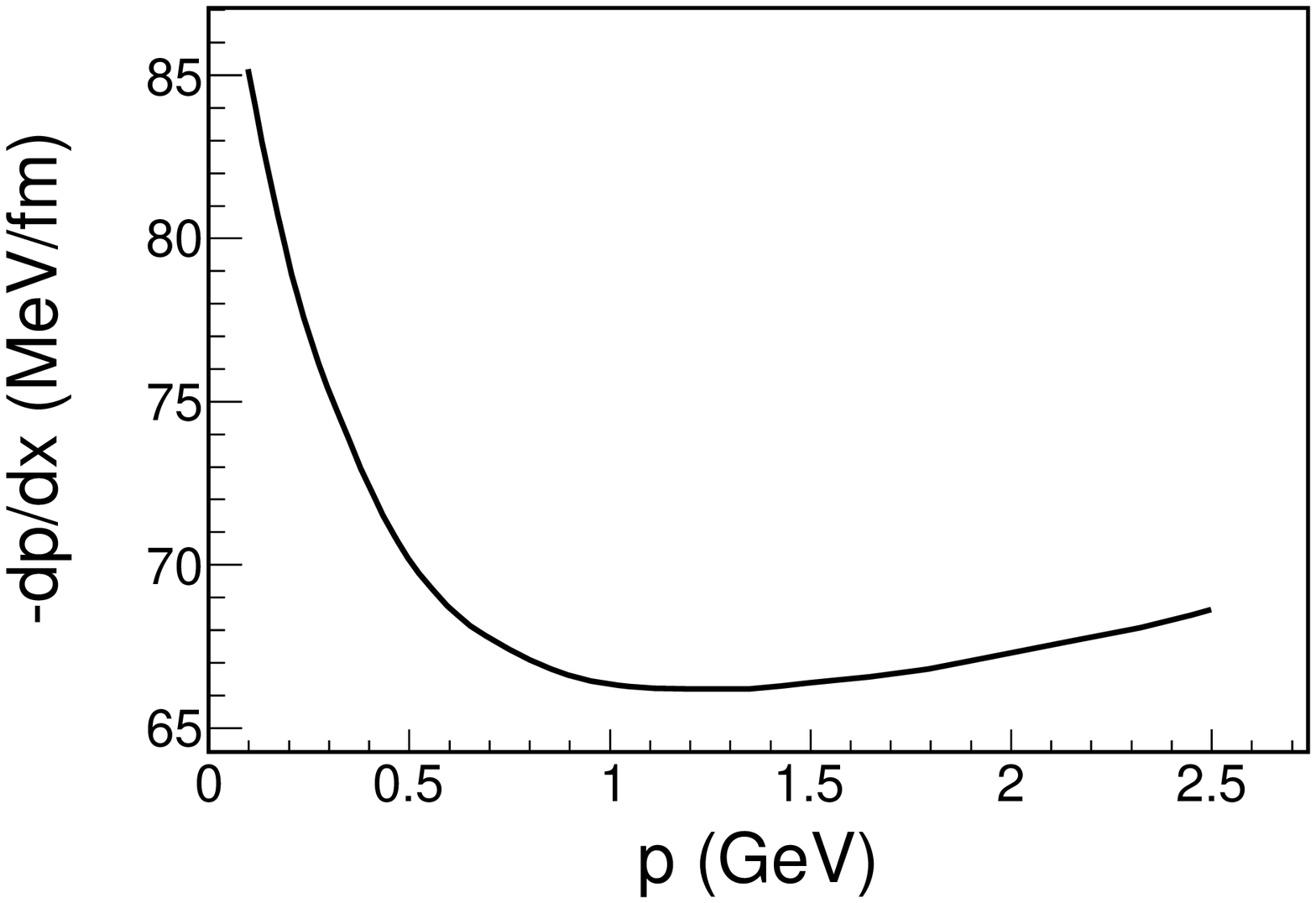}
\caption{\label{fig:energy-momentum-loss} 
Energy and momentum loss per unit length of bottomed mesons at a temperature of
150~MeV in the full meson gas consisting of pions, kaons and etas.}
\end{figure}

In the present approach we can also estimate the energy and momentum loss per
unit length of a bottomed meson traveling in the meson gas from the classical
interpretation of the $F$ coefficient as a drag force. One has
\be
dE/dx = - F\, p \, ,
\quad
\textrm{and}
\quad
dp/dx = - F\, E \, ,
\ee
in terms of the energy and momentum of the bottomed meson. These quantities are
depicted in Fig.~\ref{fig:energy-momentum-loss} at $T=150$~MeV for the $SU(3)$ meson gas.
Thus, a reference bottom quark in a $B$ or $B^*$ mesonic state with a typical
momentum of 1~GeV with respect of the rest frame of the surrounding medium,
loses about 70~MeV per Fermi as it propagates in the hadron gas. This effect
should be taken into account to correctly analyze heavy-meson distributions,
suggesting that 
the hadronic phase also has to be treated in hydrodynamical simulations
regarding open heavy flavor probes.

\begin{figure}
\centering
\includegraphics[{height=6.0cm,width=8.5cm}]{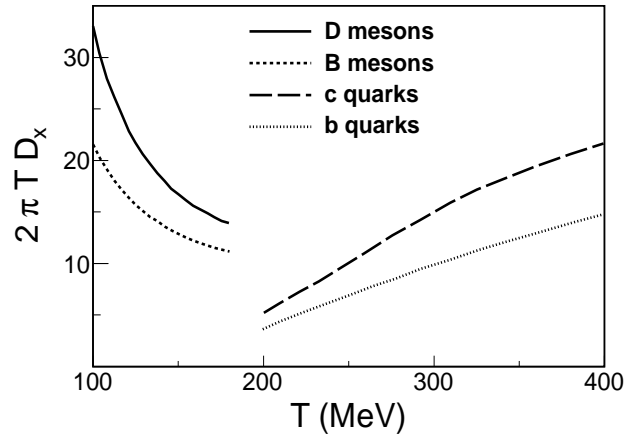}
\caption{\label{fig:Dx-comparison} 
Charm and bottom spatial diffusion coefficients below (this work) and above
(\cite{Rapp:2008qc,vanHees:2007me,He:2012df}) the crossover.}
\end{figure}

Finally, we calculate the spatial difusion coefficient, $D_x$, which can be
related to the diffusion coefficient in static limit as $D_x = T^2/\Gamma$
(cf.~\cite{LADCFLEJT:2011}, Appendix B for details). It is shown in
Fig.~\ref{fig:Dx-comparison} together with our previous result for charmed mesons, below the
crossover, and with the estimations by Rapp and van Hees for $c$- and $b$-quark
transport properties above the critical
temperature \cite{Rapp:2008qc,vanHees:2007me,He:2012df}. The conclusion observed for the charm sector is
 here confirmed
for bottom: the minimum relaxation time for heavy flavor seems to take place
around the crossover, where one expects strongest (and long-range) interactions.
Our present work reinforces the use of heavy flavor as probes of the QCD phase
transition by reducing theoretical uncertainties regarding the dynamics in the
hadronic phase.

\section{\label{sec:summary}Summary and conclusions}

In this work the drag and diffusion coefficients of bottomed mesons in a thermal
gas of pions, kaons and $\eta$ mesons have been evaluated in a Fokker-Plank transport
approach. The dynamics of the interaction of $B$ and $B^*$ with the light mesons
has been modeled employing a unitarized version of heavy-meson chiral
perturbation theory within the constraints of heavy quark symmetry. The
relevant scattering amplitudes have been calculated at next-to-leading order in
the chiral expansion and leading order in the heavy quark limit, and the free parameters have been
constrained by available data on the $B^{(*)}$ spectrum and heavy-quark considerations.
\par
We have observed a sizable temperature dependence of the transport coefficients,
indicating that both drag and diffusion are more efficient in the hotter stages
of the hadronic phase in a heavy-ion collision.
\par
Regarding momentum dependence,
it turns out that $F$ and $\Gamma_0$ change only mildly, whereas the transverse
coefficient, $\Gamma_1$, is
particularly sensitive to the momentum of the heavy meson. The latter may lead
to observable consequences in the analysis of anisotropic observables such as
the elliptic flow.
\par
In addition, a detailed comparison has been accomplished between our outcomes and the other ones in literature obtained via different approaches. Our statements
rely on good control of the elementary $B$-meson interaction with the light
meson octet over a wide range of energies, accounting for dynamical generation
of resonances in attractive channels. We have shown that preserving unitarity in the scattering amplitudes plays an essential role in providing realistic estimates of the transport coefficients at high temperatures.
\par
We have performed several consistency tests of our results such as studying the
scaling of transport coefficients with the heavy-quark mass and verifying the
Einstein relation in the static limit, which were satisfactory. 
\par
Moreover, the individual contribution of pions, kaons and $\eta$
mesons to the transport coefficients have been estimated. The most relevant contribution is led by
the pion gas, as expected, with the next-to-leading contribution being provided
by kaons and antikaons. 
\par
Also, some estimations of relevant quantities have been done directly from the transport coefficients. One example is the bottom relaxation length, which is about 80 fm for bottomed mesons with momenta of 1~GeV and for meson gas at
temperature of 150 MeV. It allows us to infer that bottomed mesons barely relax during the lifetime of the 
hadron gas, unlike charm mesons that, while not relaxing completely, lose a great deal of memory of the initial state. In this sense, the bottomed mesons constitute an optimal system to characterize the early stages of a relativistic heavy-ion
collision. Observables like the nuclear suppression factor or the elliptic flow can provide clear indications of the system properties and evolution after the nuclear collision.
\par
Furthermore, the computation of the spatial diffusion coefficient for the present case is in agreement with the idea that the relaxation time for heavy quarks has a minimum around the crossover to the quark-gluon plasma, the expected place of strongest interactions.  
\par
Another interesting quantity evaluated has been the momentum loss per unit length, which is 70~MeV per Fermi for a bottomed meson with momentum of 1~GeV propagating in the hadron gas. This result indicates that to correctly analyze the heavy-meson distributions, 
the effect of loss of energy and momentum must be taken into account.
\par
Hence, the findings of the present work discussed above reinforce the role of heavy mesons as probes of the strongly interacting matter phase
transition, paving the way to a better understanding of heavy-flavor dynamics and transport properties below the crossover.

\acknowledgments
 We want to thank Feng-Kun Guo, Juan Nieves, Santosh Ghosh, Christine Davies and Rachel Dowdall for clarifications
and comments. We especially acknowledge Felipe J. Llanes-Estrada for invaluable
discussions and suggestions. 

We thank financial support from grants
FPA2011-27853-C02-01, FPA2011-27853-C02-02,FIS2008-01323 (Ministerio de Econom\'{\i}a y Competitividad, Spain) and from the EU Integrated
Infrastructure Initiative Hadron Physics Project under Grant Agreement
n.~227431. LMA thanks CAPES (Brazil) for partial financial support. 
D.C. acknowledges financial support from Centro Nacional de
F\'isica de Part\'iculas, Astropart\'iculas y Nuclear
(CPAN, Consolider-Ingenio 2010) postdoctoral programme.
J.M.T.-R. is a recipient of an FPU grant (Ministerio de Educaci\'on, Cultura y Deporte, Spain). 

\bigskip

\appendix
\section{\label{sec:subtraction}Subtraction constant}

In paralel with the work~\cite{LADCFLEJT:2011} we would like to provide a reasonable description
of meson-meson scattering in such a way that the pertinent resonances in the different channels
are generated.
In particular, we would like to reproduce the $B_0$ and $B_1$ resonances in the $B\pi$ and $B^*\pi$
channels, respectively. However, the masses and widths of these resonances are not yet experimentally known. 

Starting at LO, there are no free low-energy constants and the perturbative
amplitude $V$ in the $B^* \pi$ channel is fixed. In the unitarized amplitude
--the one which generates a pole in the amplitude-- we introduce a subtraction
constant $a(\mu)$. This constant will be fixed by matching the function $G(s)$
in dimensional regularization to the one with cutoff regularization scheme.

The loop function $G(s)$ in dimensional regularization is given in
Eq.~(\ref{propdr}). The regularization scale $\mu$ is chosen at a natural scale
of 1~GeV \footnote{There are two typos in \cite{LADCFLEJT:2011}. The
regularization scale $\mu$ should read 1 GeV and $a(1 \textrm{ GeV})=-1.85$.}

In a cutoff regularization one has~\cite{Guo:2005wp}
\begin{widetext}
\begin{eqnarray}
 \label{eq:Gscut} G^{\Lambda}(s) &=&\frac{1}{16 \pi^2 s} \left\{ 2 i \sqrt{s} q \left( \arctan \frac{s+m_B^2-m_\pi^2}{2 i \sqrt{s} q \sqrt{1+ \frac{m_B^2}{\Lambda^2}}} +
\arctan \frac{s-m_B^2+m_\pi^2}{2 i \sqrt{s} q \sqrt{1+ \frac{m_\pi^2}{\Lambda^2}}}  \right) \right. \\ 
&-& \left. \left[ \left( s + m_B^2- m^2_\pi \right) \log \left( \frac{\Lambda}{m_B} + \sqrt{1+\frac{\Lambda^2}{m^2_B}} \right)  
+\left( s - m_B^2 + m^2_\pi \right) \log \left( \frac{\Lambda}{m_\pi} + \sqrt{1+\frac{\Lambda^2}{m^2_\pi}}   \right)  \right] \right\} \ . \nonumber 
\end{eqnarray}
\end{widetext}

\begin{figure}
\centering
\includegraphics[{height=6.0cm,width=8.5cm}]{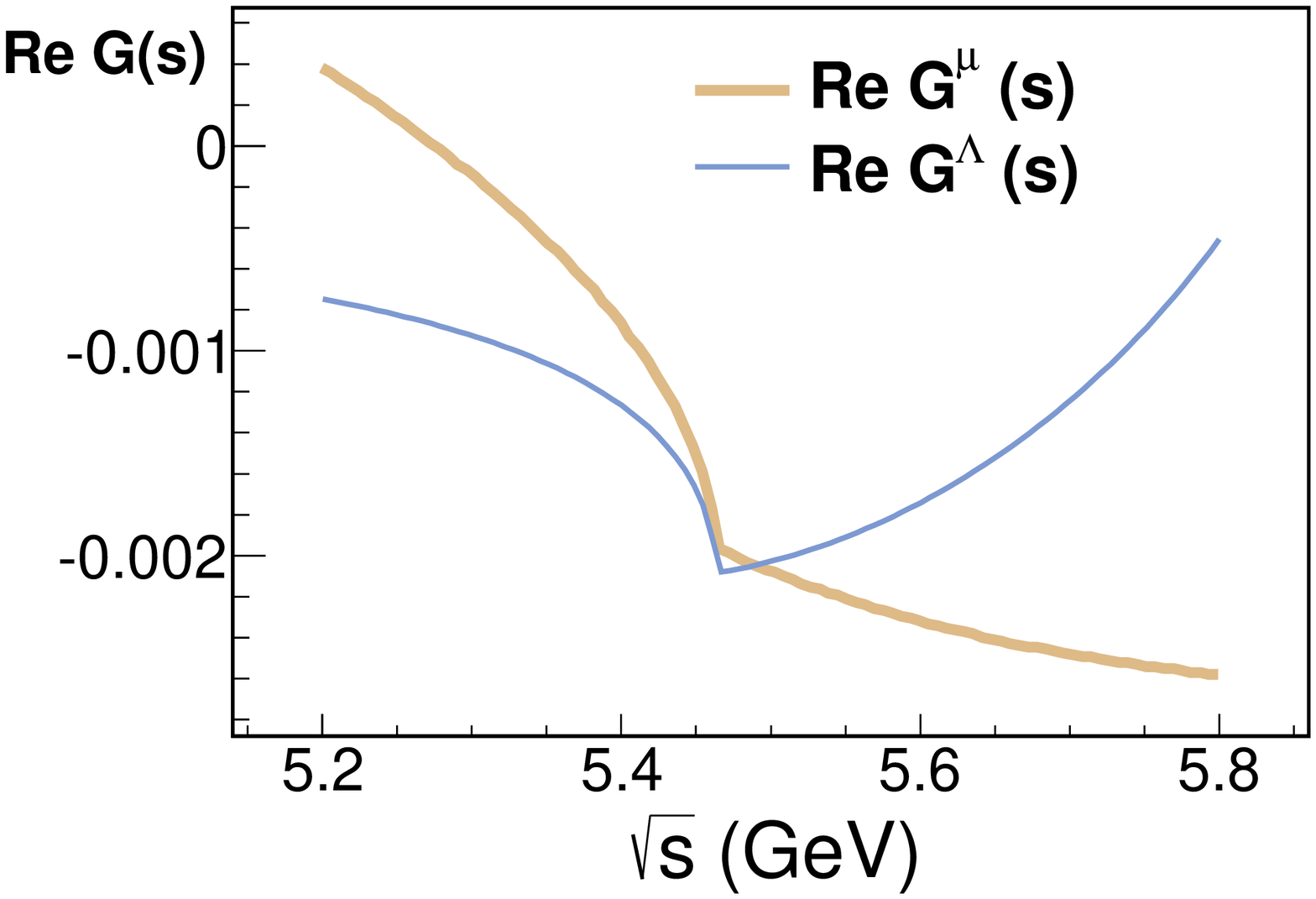}
\includegraphics[{height=6.0cm,width=8.5cm}]{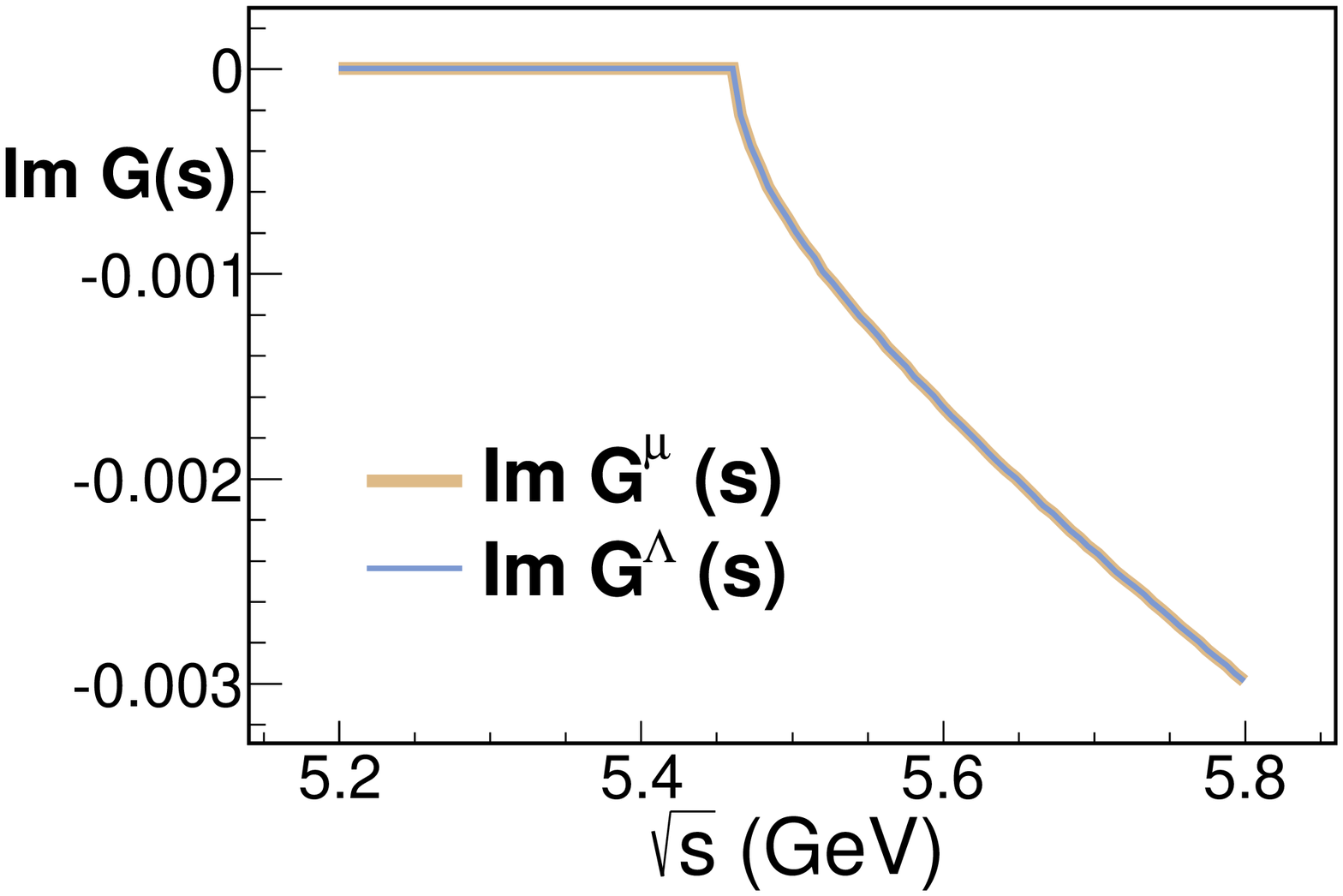}
\caption{\label{fig:loopfunc} Re $G(s)$ and Im $G(s)$ from Eq.~(\ref{propdr}) (dimensional regularization) and Eq.~(\ref{eq:Gscut}) (cutoff regularization). 
We match the loop function in the two regularization schemes at $\sqrt{s_{th}}=m_B+m_{\pi}$. A higher energies $G^{\Lambda}(s)$ becomes non-physical and
diverges around $\sqrt{s_{\infty}}=6$ GeV.
}
\end{figure}

Note that this expression is valid below and above threshold. The equivalent
cutoff in momentum,  $\Lambda$, is obtained by matching both expressions where
we use the value of $a(\mu)$ in the $D\pi$ sector \cite{LADCFLEJT:2011}.  It is
well known that $G_{\Lambda}$ presents a spurious divergence at some
$s_{\infty}$ above threshold, where its determination is
unreliable. For this reason, one typically compares the two loop
functions at threshold $s_{th}=(m_B+m_\pi)^2$ (see Fig.~\ref{fig:loopfunc}).

Combining all these requirements we obtain $a(1 \textrm{ GeV})=-3.47$,
corresponding to a reasonable cutoff momentum of about 1~GeV. 
Within this scheme we obtain a resonance around 5580~MeV at LO in the $I=1/2$
channel, about 100~MeV below the reference value of the mass of the $B_1$ (heavy
quark spin 3/2). At NLO the two free low energy
constants $h_3$ and $h_5$ can modify the pole position and width. Varying these
parameters within the
constraints of HQ Symmetry, a
maximal value of 5587~MeV can be found for the pole position, with a width around 245 MeV.
Although higher values of the resonance mass can be forced by tuning the
subtraction constant, in order to keep the equivalent cutoff of natural scale
we content ourselves with a $B_1$ pole mass of 5587~MeV and 
5534~MeV for the $B_0$ resonance (with a width of 210~MeV). Our
results agree with previous studies using similar unitarization methods.
They are summarized in Table~\ref{tab:resonances}.

\begin{widetext}
\begin{center}
\begin{table}[ht]
\begin{tabular}{|c||c|c|c|c|}
\hline 
Reference & $M (B_0)$ (MeV) & $\Gamma (B_0)$ (MeV) & $M(B_1)$ (MeV) & $\Gamma (B_1)$ (MeV) \\
\hline
This work &  5534 & 210 & 5587 & 245 \\
\cite{Guo:2006fu} & $5536\pm 29$ & $234 \pm 86$ &- & - \\
\cite{Guo:2006rp} & - & - & $5581 \pm 5$ & $220 \pm 15$ \\
\cite{Kolomeitsev:2003ac} & 5526 & - & 5590 & -\\
\cite{Flynn:2007ki} & 5600 & - & - & - \\
\cite{Gregory:2010gm},\cite{Dowdall:2012ab} & $5630 \pm 83$&- & $5693 \pm 43$ &- \\
\hline
\end{tabular}
\caption{ \label{tab:resonances} Resonance parameters obtained in different works. The last entry corresponds to our estimate from the data given in \cite{Gregory:2010gm} and \cite{Dowdall:2012ab}. In this case,
there is an uncertainty due to the coupling to decay channels which is not included in the quoted error. This uncertainty would presumably increase the error bar in more than 25 MeV.}
\end{table}
\end{center}
\end{widetext}


\end{document}